\documentclass[a4paper,onecolumn,9pt]{article}
\usepackage{graphicx}
\usepackage{hhline}
\usepackage{mathrsfs}
\usepackage{amssymb}
\usepackage{setspace}

\begin{document}
\Large\textbf{\textrm{\hspace{-1.2em} Flux-Vector-Splitting (FVS) method for Z4 formalism and its numerical analysis}}\\ \\
\large\textmd{Ryosuke Yano}\\
\small\textit{Department of Advanced Energy, University of Tokyo, 5-1-5 Kashiwanoha, Kashiwa, Chiba 277-8561, Japan \\
Email: yano@daedalus.k.u-tokyo.ac.jp}\\ \\
\large\textmd{Kojiro Suzuki}\\
\small\textit{Department of Advanced Energy, University of Tokyo, 5-1-5 Kashiwanoha, Kashiwa, Chiba 277-8561, Japan \\
Email: kjsuzuki@k.u-tokyo.ac.jp, Tel: +81-4-7136-3828,\\
Fax: +81-4-7136-3828}
\\ \\
\large\textmd{Hisayasu Kuroda}\\
\small\textit{Super Computing Division, Information Technology Center, University of Tokyo, 2-11-16 Yayoi, Bunkyo, Tokyo, Japan \\
Email: kuroda@pi.cc.u  -tokyo.ac.jp}\\ \\
\hspace{-0.0em}\textsf{\textbf{Abstract}}\\\\
The previously developed Flux-Vector-Splitting (FVS) method was formulated here for the first-order version of Z4 formalism. Then, the characteristics of this method formulated for Z4 formalism were studied by numerically analyzing the evolution of two types of black holes (free and stuffed). Finally, these numerical results from the FVS method were compared with those from the Local Lax Friedrichs and Modified Local Lax Friedrichs methods to reveal the dependency of numerical solution by Z4 formalism on the choice of numerical scheme.
\newpage
\hspace{-1.4em}\textsf{\textbf{I. Introduction}}\\\\
Numerical relativity has been developed to solve Einstein equations. Various formalisms derived from (3+1) ADM (Arnowit-Deser-Meisner) formalism \cite{Arnowit}, \cite{Shibata}, \cite{Bona3} have been proposed to solve Einstein equation robustly. Among them, Z4 formalism \cite{Bona} is constructed to have strong hyperbolicity, and as a result, techniques in CFD (Computational Fluid Dynamics) can be used to solve Z4 formalism. Bona has also reported numerical results of black hole's evolutions by Z4 formlaism with numerical schemes, such as LLF (Local Lax Friedrichs) and MLLF (Modified-Local Lax Friedrichs) methods \cite{Bona4}. In this current study, the FVS (Flux-Vector-Splitting) method recently proposed by Steger-Warming \cite{Steger} was formulated here for the first-order version of Z4 formalism. Then, numerical results from this formulated FVS method were then compared with results from the LLF and MLLF methods. The advantage of the FVS method over the LLF and MLLF methods is that the FVS method decomposes all characteristic waves of the convective matrix (the LLF and MLLF methods decompose only the maximum and minimum characteristic waves). 
Finally, the dependency of Z4 formalism on the choice of the numerical scheme was evaluated based on numerical analyses of three-dimensional evolution of two types of black holes (free \cite{Bona} and stuffed \cite{Bona}).
\\ \\
\textsf{\textbf{II. Z4 formalism}}\\\\
The first-order version of Z4 formalism is written as follows \cite{Bona2},
\begin{eqnarray}
\partial_t A_i+\partial_l\left[-\beta^l A_i+{\delta^l}_i \left(\alpha Q+\beta^m A_m \right)\right]&=&{B_i}^lA_l-{B_l}^lA_i, \\
\partial_t {B_k}^i+\partial_l\left[-\beta_l {B_k}^i+{\delta^l}_k\left(\alpha Q_i+\beta^m {B_m}^i\right)\right]&=&{B_l}^i{B_k}^l-{B_l}^l{B_k}^i,\\
\partial_t D_{kij}+\partial_l\left[-\beta_l D_{kij}+{\delta^l}_k\left(\alpha Q_{ij}+\beta^m D_{mij}\right)\right]&=&{B_k}^lD_{lij}-{B_l}^l D_{kij},\\
\partial_t K_{ij}+\partial_k \left[-\beta^k K_{ij}+\alpha {\lambda^k}_{ij} \right]&=&S\left(K_{ij}\right),\\
\partial_t \Theta+\partial_k\left[-\beta^k \Theta+\alpha\left(D^k-E^k-Z^k\right)\right]&=&S\left(\Theta\right),\\
\partial_t Z_i+\partial_k \left[-\beta^k Z_i+\alpha\left\{-{K^k}_i+{\delta^k}_i\left(trK-\Theta\right)\right\}\right]&=&S\left(Z_i\right),
\end{eqnarray}
where ${\lambda^k}_{ij}$is defined with ordering parameter $\xi$ as

\begin{eqnarray}
{\lambda^k}_{ij}&=&{D^k}_{ij}-\frac{1}{2}\left(1+\xi\right)\left({D_{ij}}^k+{D_{ji}}^k\right)+\frac{1}{2}{\delta^k}_i\left[A_j+D_j-\left(1-\xi\right)E_j-2Z_j\right] \nonumber \\
&&+\frac{1}{2}{\delta^k}_j\left[A_i+D_i-\left(1-\xi\right)E_i-2Z_i\right],
\end{eqnarray}
with the following definitions:
\begin{eqnarray}
D_i\equiv{D_{ik}}^k,\\
E_i\equiv{D^k}_{ki}.
\end{eqnarray}
In eqs. (1)-(3), $A_i$,${B_k}^i$ and $D_{kij}$ are defined as
\begin{eqnarray}
&&A_i \equiv \partial_i \ln \alpha, \\
&&{B_k}^i \equiv \partial_k \beta^i, \\
&&D_{kij} \equiv \frac{1}{2}\partial_k \gamma_{ij}.
\end{eqnarray}
In eqs. (1)-(5) and (10), $\alpha$ is called the lapse and $\beta^i$ is the shift vector.\\
In eq. (12), the intrinsic curvature $\gamma_{ij}$ is then given by
\begin{eqnarray}
\partial_t \gamma_{ij}+\partial_l \left[-\beta^l \gamma_{ij}\right]=-2 \alpha K_{ij},
\end{eqnarray}
where $K_{ij}$ is the extrinsic curvature.\\
In eqs. (4)-(6), source terms $S(K_{ij})$, $S(\Theta)$ and $S(Z_i)$ are those expressed by Bona \cite{Bona2}. \\
In eq. (1), $Q$ is the rate of temporal variation of $\alpha$ as
\begin{eqnarray}
\partial_t \alpha=-\alpha^2 Q,
\end{eqnarray}
and is expressed in harmonic coordinates as \cite{Bona2}
\begin{eqnarray}
Q=-a\frac{\beta^k}{\alpha}\partial_k \ln \alpha+f(\alpha)\left(trK-m\Theta\right).
\end{eqnarray}
In eq. (15), $f = 0$ is geodesic slicing, $f =1$ is harmonic slicing, and $f = 2/\alpha$ is $"1+\log"$ slicing. \\
In eq. (2), $Q^i$ is the rate of the temporal variation of $\beta^i$ as
\begin{eqnarray}
\partial_t \beta^i=-\alpha Q^i,
\end{eqnarray}
and $Q^i$ is expressed in harmonic coordinates as \cite{Bona2}
\begin{eqnarray}
Q^i=-\frac{\beta^k}{\alpha}\partial_k \beta^i-\alpha \gamma^{ki}\left(\partial^j\gamma_{jk}-\partial_k \ln \sqrt{\gamma}-\partial_k \ln \alpha\right).
\end{eqnarray}
\textsf{\textbf{III. Numerical scheme}}\\\\
In the numerical scheme to formulate the FSV method with Z4 formulism, we set $\beta^i=0$, because a value of 0 is typical in the evolution of a single black hole \cite{Bona}. Consequently, 31 independent variables needed to be solved: $\mathscr{U}$=($A_x$, $A_y$, $A_z$, $D_{xxx}$, $D_{xxy}$, $D_{xxz}$, $D_{xyy}$, $D_{xyz}$, $D_{xzz}$, $D_{yxx}$, $D_{yxy}$, $D_{yxz}$, $D_{yyy}$, $D_{yyz}$, $D_{yzz}$, $D_{zxx}$, $D_{zxy}$, $D_{zxz}$, $D_{zyy}$, $D_{zyz}$, $D_{zzz}$, $K_{xx}$, $K_{xy}$, $K_{xz}$, $K_{yy}$, $K_{yz}$, $K_{zz}$, $\Theta$, $Z_x$, $Z_y$, $Z_z$) are solved.\\
Equations (1), (3)-(6) are rewritten as follows by using three $31\times31$ square matrices, namely, $M$, $L$ and $N$:
\begin{eqnarray}
\partial_t \mathscr{U}+\partial_x \alpha M\mathscr{U}+\partial_y \alpha L\mathscr{U}+\partial_z \alpha N\mathscr{U}=S
\end{eqnarray}
In eq. (18), $\alpha M$ can be rewritten as follows by using the matrix $R$, which is composed of the lines of eigenvectors of $\alpha M^t$ and the diagonal matrix $\Lambda$, whose diagonal elements are eigenvalues of $\alpha M^t$: 
\begin{eqnarray}
\alpha M=R\Lambda R^{-1},
\end{eqnarray}
 where $R^{-1}$ is the inverse matrix of $R$. The formulation of $R$, $R^{-1}$ and $\Lambda$ is relatively straightforward (elements of $M$, $R$, $R^{-1}$ and $\Lambda$ in eq. (19) are shown in the Appendix). In contrast, the formulation of the eigenvectors of $\alpha{L}^t$ and $\alpha{N}^t$ is complex. To simplify this formulation, we therefore rewrote eq. (18) as follows
\begin{eqnarray}
\partial_t \mathscr{U}+\partial_x \alpha M\mathscr{U}+\psi_{y,x}\partial_y \alpha \tilde{L}\mathscr{V}+\psi_{z,x}\partial_z \alpha \tilde{N}\mathscr{W}=S,
\end{eqnarray}
where $\mathscr{V}$=($A_y$,$A_x$,$A_z$,$D_{yyy}$,$D_{yxy}$,$D_{yyz}$,$D_{yxx}$,$D_{yxz}$,$D_{yzz}$,$D_{xyy}$,$D_{xxy}$,$D_{xyz}$,$D_{xxx}$,\\
$D_{xxz}$,$D_{xzz}$,$D_{zyy}$,$D_{zxy}$,$D_{zyz}$,$D_{zxx}$,$D_{zxz}$,$D_{zzz}$,$K_{yy}$,$K_{xy}$,$K_{yz}$,$K_{xx}$,$K_{xz}$,$K_{zz}$,$\Theta$,$Z_y$,$Z_x$,$Z_z$)\\
 and $\mathscr{W}$=($A_z$,$A_y$,$A_x$,$D_{zzz}$,$D_{zyz}$,$D_{zxz}$,$D_{zyy}$,$D_{zxy}$,$D_{zxx}$,$D_{yzz}$,$D_{yyz}$,$D_{yxz}$,$D_{yyy}$,\\
$D_{yxy}$,$D_{yxx}$,$D_{xzz}$,$D_{xyz}$,$D_{xxz}$,$D_{xyy}$,$D_{xxy}$,$D_{xxx}$,$K_{zz}$,$K_{yz}$,$K_{xz}$,$K_{yy}$,$K_{xy}$,$K_{xx}$,$\Theta$,$Z_z$,$Z_y$,$Z_x$). In eq. (20), $\psi_{y,x}$ and $\psi_{z,x}$ are matrices whose elements are constants, either $0$ or $1$, and defined as
\begin{eqnarray}
\mathscr{U}\equiv \psi_{y,x}\mathscr{V} \equiv \psi_{z,x}\mathscr{W}
\end{eqnarray}
In eq. (20), we introduce $\tilde{L}$ and $\mathscr{V}$, which are obtained by exchanging the superscript and subscript $x$ for $y$ in $L$ and $\mathscr{U}$. As a result, we consider $\partial_y \alpha \tilde{L}\mathscr{V}$ in eq. (20) instead of $\partial_y \alpha L \mathscr{U}$ in eq. (18). Eigenvalues and eigenvectors of $\alpha\tilde{L}^t$ are obtained by exchanging superscript $x$ for $y$ of $\gamma^{ij}$ in each element of $R$ and $\Lambda$ shown in the Appendix. The same exchange of superscripts is done for $R^{-1}$ to obtain the inverse matrix composed of the lines of eigenvectors of $\alpha\tilde{L}^t$. As with $\alpha N \mathscr{U}$, we exchange the superscript and subscript $x$ for $z$ in $N$ and $\mathscr{U}$, thus yielding $\alpha \tilde{N} \mathscr{W}$. The eigenvalues and eigenvectors of the matrix $\tilde{N}^t$ are obtained by exchanging the superscript $x$ for $z$ of $\gamma^{ij}$ in each element of $R$ and $\Lambda$. The FVS method discretizes $\partial_x \alpha M \mathscr{U}$ in eq. (18) as \cite{Steger}
\begin{eqnarray}
&& \left(\partial_x \alpha M \mathscr{U}\right)_i\simeq \frac{\mathscr{F}_{i+1/2}-\mathscr{F}_{i-1/2}}{\Delta x_i}, \\
&&\mathscr{F}_{i+1/2}=\frac{1}{2}\left\{\mathscr{F}_{i+1}+\mathscr{F}_{i}-\left(G_{i+1}-G_i\right)\right\},\\
&& \mathscr{F}_{i} \equiv  \alpha_{i}M_{i}\mathscr{U}_{i},\\
&& G_i \equiv R_{i}|\Lambda_{i}|R_{i}^{-1}\mathscr{U}_{i},
\end{eqnarray}
where $i$ corresponds to the $i^{th}$ numerical grid on the x-axis and $\Delta x_i$ is the grid size given by $\Delta x_i=|x_{i+1}-x_i|$. Similar to the descretization of eq. (18) shown in eqs. (22)-(25), descretization is done for $\partial_y \alpha \tilde{L}\mathscr{V}$ and $\partial_z\alpha\tilde{N}\mathscr{W}$ in eq. (20).\\
In contrast to the FVS method, the HLL (Harten-Lee-vanLeer) method \cite{Toro} requires only eigenvalues without eigenvectors. The numerical flux $\mathscr{F}_{i+1/2}$ in eq. (22) is defined by the HLL method as
\begin{eqnarray}
\mathscr{F}_{i+1/2}=\frac{s_{R,i+1/2}\mathscr{F}_i-s_{L,i+1/2}\mathscr{F}_{i+1}+s_{L,i+1/2}s_{R,i+1/2}\left(\mathscr{U}_{i+1}-\mathscr{U}_i\right)}{s_{R,i+1/2}-s_{L,i+1/2}},
\end{eqnarray}
where $R=\mbox{Right}$ and $L=\mbox{Left}$, and $s_{R,i+1/2} \ge 0$ and $s_{L+1/2} \le 0$ in eq. (26) are defined as
\begin{eqnarray}
s_{R,i+1/2}=\max\left(\Lambda_{i+1/2}\right).\\
s_{L,i+1/2}=\min\left(\Lambda_{i+1/2}\right).
\end{eqnarray}
In Z4 formalism with $\beta_i=0$, $|s^x_R|=|s^x_L|$. Consequently, in the case of zero shift (i.e., $\beta_i=0$), the HLL method is the same as the LLF(Local Lax-Friedrichs) method.\\
In the modified Local Lax Friedrichs (MLLF) method proposed by \cite{Bona4}, the numerical flux in eq. (26) is replaced as follows:
\begin{eqnarray}
\mathscr{F}_{i+1/2}=\frac{1}{2}\left\{\mathscr{F}_i+\mathscr{F}_{i+1}-\left(s_{R,i+1} \mathscr{U}_{i+1}+s_{L,i} \mathscr{U}_i\right)\right\}.
\end{eqnarray} 
Both the LLF method and MLLF method use the same descretization procedure for $\partial_y\alpha L \mathscr{U}$ and $\partial_z \alpha N \mathscr{U}$ in eq. (18).
Comparison of eq. (23) with eqs. (26) and (29) reveals that the FVS method only decomposes all of the characteristic waves of the convective matrix $\alpha M$, whereas the LLF and MLLF methods decompose only the maximum and minimum characteristic waves. Consequently, the highest accuracy is expected when all the waves are considered. The FVS, LLF and MLLF methods considered in this study are first-order schemes. Higher order schemes can be formulated by applying the MUSCL (Monotone Upstream-centered Schemes for Conservation Laws) method \cite{MUSCL} to all schemes. In this study, however, we analyzed only the spatial and temporal first-order schemes to focus on the choice of numerical scheme itself.
\\\\
\textsf{\textbf{IV. Initial conditions for the evolution of a single black hole}}\\\\
In isotropic coordinates, the initial conditions for the evolution of a single black hole are as follows \cite{Bona}
\begin{eqnarray}
&&K_{ij}|_{t=0}=0, \  \  \  \gamma_{ij}|_{t=0}=\Psi^4 \delta_{ij}, \\
&&\alpha|_{t=0}=1, \  \  \  \beta^i|_{t=0}=0, \\
&&\Theta|_{t=0}=0, \  \  \  Z_i|_{t=0}=0. 
\end{eqnarray}
In eq. (30), $\Psi^4$ is defined as \cite{Bona}
\begin{eqnarray}
\Psi^4&=&\left(1+\frac{M}{\rho}\right) \  \  \  \  \  \  \  \  \   \   \   \  \   \  \rho \ge \frac{M}{2}\nonumber \\
&=& 64 \left[1+\left(\frac{2\rho}{M}\right)^2 \right]^{-2} \  \  \ \rho < \frac{M}{2},
\end{eqnarray}
where $\rho$ is the distance from the center of the black hole (i.e., the origin in isotropic coordinates), and $M$ is the mass of the black hole. In eq. (15), the parameter $f$ for the black hole's evolutions is set to "1+log" slicing as
\begin{eqnarray}
f&=&\frac{2}{\alpha}.
\end{eqnarray}
In eq. (7), $\xi$ for the evolutions of single black hole is set to
\begin{eqnarray}
\xi=0.
\end{eqnarray}
In this study, we considered two types of black holes: a free black hole and a stuffed black hole. 
The difference between these two types is their energy density ($\tau$) distribution inside the Schwartzchild radius ($\rho=M/2$) as follows.
\begin{eqnarray}
&&\tau=0 \  \  \  \  \  \mbox{(Free black Hole)}\\
&&\tau=\frac{3}{4M^2} \  \  (\rho < \frac{M}{2}) \ \ \  \ \tau=0 \  \  \ (\rho \ge M/2) \  \  \  \mbox{(Stuffed black hole)}
\end{eqnarray}
\\ \\
\textsf{\textbf{V. Numerical analysis of the evolution of a single black hole}}\\\\
In this section, the evolution of the two types of single black holes is numerically analyzed using the FVS, LLF and MLLF methods. The numerical grid is a $101 \times 101 \times 101$ unequally sized grid in Cartesian coordinates $-50M \le X,Y,Z \le 50M$ as shown in Fig. 1. The grid is denser near the origin (i.e., center of the black hole). First, we analyzed the evolution of a single free black hole using the same $m$ values in eq. (15) reported by Bona, namely, $m=0$ and $m=-3$. \\ \\
\underline{\textsf{Evolution of a free black hole for $m=0$}}\\ \\
Figures 2A, B and C show the lapse profiles along the $x$-axis in increments of $\Delta t/M=0.5$ using the FVS, LLF and MLLF methods, respectively.
Similar to the result obtained by Bona \cite{Bona}, the lapse did not stabilize, regardless of the method used. However, the time that the lapse became unstable was later for the FVS method than for either the LLF or MLLF methods, and the dynamics of this instability differed for the FVS method. In summary, the lapse by FVS method goes forward by the blow-up, although lapses by LLF and MLLF methods never go forward since the reverse wave begins to propagate into the backward.\\
The validity of the code was verified by comparing our numerical results obtained using the FVS method with those obtained using the MMC (Marquina solver \cite{Marquina} with Monotonic Centered slope limiter) method by Bona \cite{Bona2}. Figure 3 shows the lapse profiles along the $x$-axis at $t/M=0, 0.5, 1.5, 2$ and $t/M=14$. In general, the lapse at $0 \le t/M \le 2$ obtained using the FVS method is similar to that obtained using the MMC method. The difference near the origin at $t/M=0.5,1.0$ between the FVS and MMC method is caused by the rough grid \cite{Bona2} used in the MMC method. The differences in results between two methods were significant at $t/M=14$. These results reveal that the dynamics of the lapse depends on the choice of the numerical scheme.\\
Figures 4 and 5 show profiles of $Z_x$ and $\Theta$, respectively, along the $x$-axis in increments of $\Delta t/M=0.5$. $Z_x$ is antisymmetric on both sides of the origin, whereas $\Theta=Z_t$ is symmetric. Figure 6 shows $trK$ profiles in increments of $\Delta t/M=0.5$. From eq. (15), the instability of the lapse can be explained by the growth of the domain $trK<0$ in Fig. 6.\\
According to numerical results obtained using the MMC method reported by Bona \cite{Bona}, the lapse will always become unstable when $m=-3$ in eq. (15). We therefore analyzed the evolution of a free black hole for $m=-3$ in eq. (15).\\ \\
\underline{\textsf{Evolution of a free black hole for $m=-3$}}\\ \\
Figure 7 shows lapse profiles along the $x$-axis in increments of $\Delta t/M=0.5$ obtained using the FVS and MLLF methods. The lapse obtained using the FVS method became unstable when $t/M$ exceeded $3.0$, whereas the lapse obtained using the MLLF method remained stable. At $t/M=3.0$, the lapse goes backward and overshoots unity remarkably. These dynamics of the instability of the lapse obtained using the FVS method is caused exclusively by the $trK+3\Theta$ term in eq. (15). \\ \\
\underline{\textsf{Evolution of a stuffed black hole for $m=0$}}\\ \\
Finally, we consider the evolution of a single stuffed black hole for $m=0$ in eq. (15). Figure 8 shows lapse profiles along the $x$-axis in increments of $\Delta t/M=0.5$. The lapse obtained using the FVS method became unstable when $t/M$ exceeded $5.5$, whereas the lapse obtained using the MLLF method still goes forward. The reverse wave going backwards emerged at $t/M=5.0$, which is earlier than the time of the emergence of the reverse wave in the evolution of the free black hole (see preceding section). Furthermore, the "serrated" profiles obtained using the MLLF method around $x/M=0.5$, which is the Schwartzchild radius, were not observed in profiles obtained using the FVS method.\\ \\
In summary, the differences between the FVS method and the LLF and MLLF methods are significant in the analysis of the evolution of free or stuffed black holes. In the evolution of a free black hole, the value of $m$ in eq. (15) to avoid instability in the lapse obtained using the MLLF method causes the lapse obtained using the FVS method to become unstable earlier. Therefore, in the evolution of either a free or stuffed black hole, suitable values of $f$ and $m$ in eq. (15) are needed for the FVS method to avoid instability of the lapse. \\ \\
\textsf{\textbf{VI.Conclusions}}\\\\
In this study, we formulated the FVS method for the first-order version of Z4 formalism, and analyzed its characteristics by solving the evolution of free and stuffed black holes. The numerical results were then compared with those from the LLF and MLLF methods, revealing that (a) the dynamics of the lapse in the evolution of free and stuffed black holes by the FVS method significantly differ from those by either the LLF or MLLF method and (b) a suitable variable rate of the lapse must be determined for the chosen numerical scheme such that the lapse does not become unstable.\\ \\
\textsf{\textbf{Acknowledgement}}\\\\
\textit{We gratefully acknowledge Dr. Carles Bona (Departament de Fisica, Universitat de les Illes Balears, Palma de Mallorca, Spain) for kind comments on the numerics of Z4 formalism. All calculations were executed using the supercomputer HITACHI SR11000 at the University of Tokyo through collaborative research with the Information Technology Center at the University of Tokyo.}
\newpage
\hspace{-1.4em}\textsf{\textbf{APPENDIX: Elements of Matrix}}, $\bf{M}$, $\bf{R}$, $\bf{R^{-1}}$ \textsf{\textbf{and}} $\bf{\Lambda}$ \textsf{\textbf{in eq. (19) for}} $\bf{\xi}=0$ \textsf{\textbf{in eq. (7)}}\\\\
Non-zero elemnts of matrix $M_{ij}$, ($0\le i,j \le 30$) are
\begin{eqnarray*}
&&M_{0,21}=f\gamma^{xx},M_{0,22}=2f\gamma^{xy},M_{0,23}=2 f\gamma^{xz},M_{0,24}=f\gamma^{yy},M_{0,25}=2f\gamma^{yz},\\
&&M_{0,26}=f\gamma^{zz},M_{0,27}=-fm, M_{3,21}=1,M_{4,22}=1,M_{5,23}=1,M_{6,24}=1,M_{7,25}=1,M_{8,26}=1,\\
&&M_{21,0}=1,M_{21,6}=\gamma^{yy},M_{21,7}=2\gamma^{yz},M_{21,8}=\gamma^{zz},M_{21,10}=-\gamma^{yy},\\
&&M_{21,11}=-\gamma^{yz},M_{21,16}=-\gamma^{yz},M_{21,17}=-\gamma^{zz},M_{21,28}=-2,M_{22,1}=\frac{1}{2},\\
&&M_{22,6}=-\gamma^{xy},M_{22,7}=-\gamma^{xz},M_{22,10}=\gamma^{xy},M_{22,11}=\frac{1}{2}\gamma^{xz},M_{22,13}=\frac{1}{2}\gamma^{yz},\\
&&M_{22,14}=\frac{1}{2}\gamma^{zz},M_{22,16}=\frac{1}{2}\gamma^{xz},M_{22,18}=-\frac{1}{2}\gamma^{yz},M_{22,19}=-\frac{1}{2}\gamma^{zz},M_{22,29}=-1,\\
&&M_{23,2}=\frac{1}{2},M_{23,7}=-\gamma^{xy},M_{23,8}=-\gamma^{xz},M_{23,11}=\frac{1}{2}\gamma^{xy},M_{23,13}=-\frac{1}{2}\gamma^{yy},M_{23,14}=-\frac{1}{2}\gamma^{yz},\\
&&M_{23,16}=\frac{1}{2}\gamma^{xy},M_{23,17}=\gamma^{xz},M_{23,18}=\frac{1}{2}\gamma^{yy},M_{23,19}=\frac{1}{2}\gamma^{yz},M_{23,30}=-1,\\
&&M_{24,6}=\gamma^{xx},M_{24,10}=-\gamma^{xx},M_{24,13}=-\gamma^{xz},M_{24,18}=\gamma^{xz},M_{25,7}=\gamma^{xx},\\
&&M_{25,11}=-\frac{1}{2}\gamma^{xx},M_{25,13}=\frac{1}{2}\gamma^{xy},M_{25,14}=-\frac{1}{2}\gamma^{xz},M_{25,16}=-\frac{1}{2}\gamma^{xx},M_{25,18}=-\frac{1}{2}\gamma^{xy},\\
&&M_{25,19}=\frac{1}{2}\gamma^{xz},M_{26,8}=\gamma^{xx},M_{26,14}=\gamma^{xy},M_{26,17}=-\gamma^{xx},M_{26,19}=-\gamma^{xy},\\
&&M_{27,6}=-\gamma^{xy}\gamma^{xy}+\gamma^{xx}\gamma^{yy},M_{27,7}=-2\gamma^{xy}\gamma^{xz}+2\gamma^{xx}\gamma^{yz},M_{27,8}=-\gamma^{xz}\gamma^{xz}+\gamma^{xx}\gamma^{zz},\\
&&M_{27,10}=\gamma^{xy}\gamma^{xy}-\gamma^{xx}\gamma^{yy},M_{27,11}=\gamma^{xy}\gamma^{xz}-\gamma^{xx}\gamma^{yz},M_{27,13}=-\gamma^{xz}\gamma^{yy}+\gamma^{xy}\gamma^{yz},\\
&&M_{27,14}=-\gamma^{xz}\gamma^{yz}+\gamma^{xy}\gamma^{zz},M_{27,16}=\gamma^{xy}\gamma^{xz}-\gamma^{xx}\gamma^{yz},M_{27,17}=\gamma^{xz}\gamma^{xz}-\gamma^{xx}\gamma^{zz},\\
&&M_{27,18}=\gamma^{xz}\gamma^{yy}-\gamma^{xy}\gamma^{yz},M_{27,19}=\gamma^{xz}\gamma^{yz}-\gamma^{xy}\gamma^{zz},M_{27,28}=-\gamma^{xx},\\
&&M_{27,29}=-\gamma^{xy},M_{27,30}=-\gamma^{xz},M_{28,22}=\gamma^{xy},M_{28,23}=\gamma^{xz},M_{28,24}=\gamma^{yy},\\
&&M_{28,25}=2\gamma^{yz},M_{28,26}=\gamma^{zz},M_{28,27}=-1,M_{29,22}=-\gamma^{xx},M_{29,24}=-\gamma^{xy},\\
&&M_{29,25}=-\gamma^{xz},M_{30,23}=-\gamma^{xx},M_{30,25}=-\gamma^{xy},M_{30,26}=-\gamma^{xz}.
\end{eqnarray*}
Non-zero elemnts of matrix $R_{ij}$, ($0\le i,j \le 30$) are
\begin{eqnarray*}
&&R_{0,0}=\frac{-2\gamma^{xz}}{\gamma^{xx}},R_{0,1}=\frac{-2\gamma^{xy}}{\gamma^{xx}},R_{0,3}=\frac{\gamma^{xz}\gamma^{yz}-\gamma^{xy}\gamma^{zz}}{\gamma^{xx}},R_{0,4}=\frac{\gamma^{xz}\gamma^{yy}-\gamma^{xy}\gamma^{yz}}{\gamma^{xx}},\\
&&R_{0,8}=\frac{-\gamma^{xz}\gamma^{yz}+\gamma^{xy}\gamma^{zz}}{\gamma^{xx}},R_{0,9}=\frac{-\gamma^{xz}\gamma^{yy}+\gamma^{xy}\gamma^{yz}}{\gamma^{xx}},\\
&&R_{0,20}=-\frac{f(-2+m)}{(-1+f)\sqrt{\gamma^{xx}}},R_{0,26}=\frac{f(-2+m)}{(-1+f)\sqrt{\gamma^{xx}}},R_{0,29}=-\sqrt{f\gamma^{xx}},R_{0,30}=\sqrt{f\gamma^{xx}},\\
&&R_{1,1}=2,R_{1,3}=-\frac{\gamma^{xz}\gamma^{xz}}{\gamma^{xx}}+\gamma^{zz},R_{1,4}=-\frac{\gamma^{xy}\gamma^{xz}}{\gamma^{xx}}+\gamma^{yz},R_{1,8}=\frac{\gamma^{xz}\gamma^{xz}}{\gamma^{xx}}-\gamma^{zz},R_{1,9}=\frac{\gamma^{xy}\gamma^{xz}}{\gamma^{xx}}-\gamma^{yz},\\
&&R_{2,0}=2,R_{2,3}=\frac{\gamma^{xy}\gamma^{xz}}{\gamma^{xx}}-\gamma^{yz},R_{2,4}=\frac{\gamma^{xy}\gamma^{xy}}{\gamma^{xx}}-\gamma^{yy},R_{2,8}=-\frac{\gamma^{xy}\gamma^{xz}}{\gamma^{xx}}+\gamma^{yz},R_{2,9}=-\frac{\gamma^{xy}\gamma^{xy}}{\gamma^{xx}}+\gamma^{yy},\\
&&R_{3,16}=1,R_{3,17}=\frac{\gamma^{xz}\gamma^{yy}}{-\gamma^{xx}(\gamma^{xy})^2+(\gamma^{xx})^2\gamma^{yy}},R_{3,18}=\frac{\gamma^{xy}\gamma^{yy}}{-\gamma^{xx}(\gamma^{xy})^2+(\gamma^{xx})^2\gamma^{yy}},\\
&&R_{3,19}=\frac{-1+(\gamma^{xy})^2/(-(\gamma^{xy})^2+\gamma^{xx}\gamma^{yy})}{\gamma^{xx}},\\
&&R_{3,20}=\frac{f(\gamma^{xy})^2(m-2)+\gamma^{xx}\gamma^{yy}\left\{1+f(1-m)\right\}}{(f-1)(\gamma^{xx}\sqrt{\gamma^{xx}})(-(\gamma^{xy})^2+\gamma^{xx}\gamma^{yy})},R_{3,21}=\frac{-(\gamma^{xz})^2\gamma^{yy}+(\gamma^{xy})^2\gamma^{zz}}{\gamma^{xx}\sqrt{\gamma^{xx}}(-(\gamma^{xy})^2+\gamma^{xx}\gamma^{yy})},\\
&&R_{3,22}=\frac{2\gamma^{xy}(-\gamma^{xz}\gamma^{yy}+\gamma^{xy}\gamma^{yz})}{\gamma^{xx}\sqrt{\gamma^{xx}}(-(\gamma^{xy})^2+\gamma^{xx}\gamma^{yy})},R_{3,23}=\frac{\gamma^{xz}\gamma^{yy}}{-\gamma^{xx}(\gamma^{xy})^2+(\gamma^{xx})^2\gamma^{yy}},\\
&&R_{3,24}=\frac{\gamma^{xy}\gamma^{yy}}{-\gamma^{xx}(\gamma^{xy})^2+(\gamma^{xx})^2\gamma^{yy}},R_{3,25}=\frac{-1+(\gamma^{xy})^2/(-(\gamma^{xy})^2+\gamma^{xx}\gamma^{yy})}{\gamma^{xx}},\\
&&R_{3,26}=\frac{(\gamma^{xy})^2/(-(\gamma^{xy})^2+\gamma^{xx}\gamma^{yy})+(-1+f(-1+m))/(-1+f)}{\gamma^{xx}\sqrt{\gamma^{xx}}},\\
&&R_{3,27}=\frac{(\gamma^{xz})^2\gamma^{yy}-(\gamma^{xy})^2\gamma^{zz}}{\gamma^{xx}\sqrt{\gamma^{xx}}(-(\gamma^{xy})^2+\gamma^{xx}\gamma^{yy})},R_{3,28}=\frac{2\gamma^{xy}(\gamma^{xz}\gamma^{yy}-\gamma^{xy}\gamma^{yz})}{\gamma^{xx}\sqrt{\gamma^{xx}}(-(\gamma^{xy})^2+\gamma^{xx}\gamma^{yy})},\\
&&R_{3,29}=-\frac{1}{\sqrt{f\gamma^{xx}}},R_{3,30}=\frac{1}{\sqrt{f\gamma^{xx}}},\\
&&R_{4,15}=1,R_{4,17}=\frac{\gamma^{xy}\gamma^{xz}}{\gamma^{xx}((\gamma^{xy})^2-\gamma^{xx}\gamma^{yy})},R_{4,18}=\frac{\gamma^{yy}}{(\gamma^{xy})^2-\gamma^{xx}\gamma^{yy}},\\
&&R_{4,19}=\frac{\gamma^{xy}}{(\gamma^{xy})^2-\gamma^{xx}\gamma^{yy}},R_{4,20}=\frac{\gamma^{xy}}{\sqrt{\gamma^{xx}}(-(\gamma^{xy})^2+\gamma^{xx}\gamma^{yy})},\\
&&R_{4,21}=\frac{\gamma^{xy}((\gamma^{xz})^2-\gamma^{xx}\gamma^{zz})}{\gamma^{xx}\sqrt{\gamma^{xx}}(-(\gamma^{xy})^2+\gamma^{xx}\gamma^{yy})},R_{4,22}=\frac{\gamma^{xz}((\gamma^{xy})^2+\gamma^{xx}\gamma^{yy})-2\gamma^{xx}\gamma^{xy}\gamma^{yz}}{\gamma^{xx}\sqrt{\gamma^{xx}}(-(\gamma^{xy})^2+\gamma^{xx}\gamma^{yy})},\\
&&R_{4,23}=\frac{(\gamma^{xy})^2}{\gamma^{xx}((\gamma^{xy})^2-\gamma^{xx}\gamma^{yy})},R_{4,24}=\frac{\gamma^{yy}}{(\gamma^{xy})^2-\gamma^{xx}\gamma^{yy}},\\
&&R_{4,25}=\frac{\gamma^{xy}}{(\gamma^{xy})^2-\gamma^{xx}\gamma^{yy}},R_{4,26}=\frac{\gamma^{xy}}{\sqrt{\gamma^{xx}}((\gamma^{xy})^2-\gamma^{xx}\gamma^{yy})},\\
&&R_{4,27}=\frac{\gamma^{xy}(-(\gamma^{xz})^2+\gamma^{xx}\gamma^{zz})}{\gamma^{xx}\sqrt{\gamma^{xx}}(-(\gamma^{xy})^2+\gamma^{xx}\gamma^{yy})},R_{4,28}=\frac{-\gamma^{xz}((\gamma^{xy})^2+\gamma^{xx}\gamma^{yy})+2\gamma^{xx}\gamma^{xy}\gamma^{yz}}{\gamma^{xx}\sqrt{\gamma^{xx}}(-(\gamma^{xy})^2+\gamma^{xx}\gamma^{yy})},\\
&&R_{5,14}=1,R_{5,17}=-\frac{1}{\gamma^{xx}},R_{5,21}=\frac{\gamma^{xz}}{\gamma^{xx}\sqrt{\gamma^{xx}}},\\
&&R_{5,22}=\frac{\gamma^{xy}}{\gamma^{xx}\sqrt{\gamma^{xx}}},R_{5,23}=-\frac{1}{\gamma^{xx}},R_{5,27}=-\frac{\gamma^{xz}}{\gamma^{xx}\sqrt{\gamma^{xx}}},R_{5,28}=-\frac{\gamma^{xy}}{\gamma^{xx}\sqrt{\gamma^{xx}}},\\
&&R_{6,4}=-\frac{\gamma^{xz}}{\gamma^{xx}},R_{6,9}=\frac{\gamma^{xz}}{\gamma^{xx}},R_{6,12}=1,\\
&&R_{6,17}=\frac{\gamma^{xz}}{-(\gamma^{xy})^2+\gamma^{xx}\gamma^{yy}},R_{6,18}=\frac{\gamma^{xy}}{-(\gamma^{xy})^2+\gamma^{xx}\gamma^{yy}},R_{6,19}=\frac{\gamma^{xx}}{-(\gamma^{xy})^2+\gamma^{xx}\gamma^{yy}},\\
&&R_{6,20}=\frac{\sqrt{\gamma^{xx}}}{(\gamma^{xy})^2-\gamma^{xx}\gamma^{yy}},R_{6,21}=\frac{(\gamma^{xz})^2-\gamma^{xx}\gamma^{zz}}{\sqrt{\gamma^{xx}}((\gamma^{xy})^2-\gamma^{xx}\gamma^{yy})},R_{6,22}=\frac{2(\gamma^{xy}\gamma^{xz}-\gamma^{xx}\gamma^{yz})}{\sqrt{\gamma^{xx}}((\gamma^{xy})^2-\gamma^{xx}\gamma^{yy})},\\
&&R_{6,23}=\frac{\gamma^{xz}}{-(\gamma^{xy})^2+\gamma^{xx}\gamma^{yy}},R_{6,24}=\frac{\gamma^{xy}}{-(\gamma^{xy})^2+\gamma^{xx}\gamma^{yy}},R_{6,25}=\frac{\gamma^{xx}}{-(\gamma^{xy})^2+\gamma^{xx}\gamma^{yy}},\\
&&R_{6,26}=\frac{\sqrt{\gamma^{xx}}}{-(\gamma^{xy})^2+\gamma^{xx}\gamma^{yy}},R_{6,27}=\frac{(\gamma^{xz})^2-\gamma^{xx}\gamma^{zz}}{\sqrt{\gamma^{xx}}(-(\gamma^{xy})^2+\gamma^{xx}\gamma^{yy})},R_{6,28}=\frac{-2(\gamma^{xy}\gamma^{xz}-\gamma^{xx}\gamma^{yz})}{\sqrt{\gamma^{xx}}((\gamma^{xy})^2-\gamma^{xx}\gamma^{yy})},\\
&&R_{7,3}=-\frac{\gamma^{xz}}{2\gamma^{xx}},R_{7,4}=\frac{\gamma^{xy}}{2\gamma^{xx}},R_{7,8}=\frac{\gamma^{xz}}{2\gamma^{xx}},\\
&&R_{7,9}=-\frac{\gamma^{xy}}{2\gamma^{xx}},R_{7,11}=\frac{1}{2},R_{7,22}=-\frac{1}{\sqrt{\gamma^{xx}}},R_{7,28}=\frac{1}{\sqrt{\gamma^{xx}}},\\
&&R_{8,3}=\frac{\gamma^{xy}}{\gamma^{xx}},R_{8,5}=1,R_{8,8}=-\frac{\gamma^{xy}}{\gamma^{xx}},R_{8,21}=-\frac{1}{\sqrt{\gamma^{xx}}},R_{8,27}=\frac{1}{\sqrt{\gamma^{xx}}},\\
&&R_{9,13}=1,R_{10,12}=1,R_{11,11}=1,R_{12,10}=1,R_{13,9}=1,R_{14,8}=1,\\
&&R_{15,7}=1,R_{16,6}=1,R_{17,5}=1,R_{18,4}=1,R_{19,3}=1,R_{20,2}=1,\\
&&R_{21,17}=\frac{\gamma^{xz}\gamma^{yy}}{\sqrt{\gamma^{xx}}((\gamma^{xy})^2-\gamma^{xx}\gamma^{yy})},R_{21,18}=\frac{\gamma^{xy}\gamma^{yy}}{\sqrt{\gamma^{xx}}((\gamma^{xy})^2-\gamma^{xx}\gamma^{yy})},\\
&&R_{21,19}=\frac{1+(\gamma^{xy})^2/((\gamma^{xy})^2-\gamma^{xx}\gamma^{yy})}{\sqrt{\gamma^{xx}}},\\
&&R_{21,20}=\frac{(\gamma^{xy})^2/(-(\gamma^{xy})^2+\gamma^{xx}\gamma^{yy})+(-1+f(-1+m))/(-1+f)}{\gamma^{xx}},\\
&&R_{21,21}=\frac{(\gamma^{xz})^2\gamma^{yy}-(\gamma^{xy})^2\gamma^{zz}}{-\gamma^{xx}(\gamma^{xy})^2+(\gamma^{xx})^2\gamma^{yy}},R_{21,22}=\frac{2\gamma^{xy}(\gamma^{xz}\gamma^{yy}-\gamma^{xy}\gamma^{yz})}{\gamma^{xx}(-(\gamma^{xy})^2+\gamma^{xx}\gamma^{yy})},\\
&&R_{21,23}=\frac{\gamma^{xz}\gamma^{yy}}{\sqrt{\gamma^{xx}}(-(\gamma^{xy})^2+\gamma^{xx}\gamma^{yy})},R_{21,24}=\frac{\gamma^{xy}\gamma^{yy}}{\sqrt{\gamma^{xx}}(-(\gamma^{xy})^2+\gamma^{xx}\gamma^{yy})},\\
&&R_{21,25}=\frac{2(\gamma^{xy}\gamma^{xy})^2-\gamma^{xx}\gamma^{yy}}{\sqrt{\gamma^{xx}}(-(\gamma^{xy})^2+\gamma^{xx}\gamma^{yy})},\\
&&R_{21,26}=\frac{(\gamma^{xy})^2/(-(\gamma^{xy})^2+\gamma^{xx}\gamma^{yy})+(-1+f(-1+m))/(-1+f)}{\gamma^{xx}},\\
&&R_{21,27}=\frac{(\gamma^{xz})^2\gamma^{yy}-(\gamma^{xy})^2\gamma^{zz}}{-\gamma^{xx}(\gamma^{xy})^2+(\gamma^{xx})^2\gamma^{yy}},R_{21,28}=\frac{2\gamma^{xy}(\gamma^{xz}\gamma^{yy}-\gamma^{xy}\gamma^{yz})}{\gamma^{xx}\left\{-(\gamma^{xy})^2+\gamma^{xx}\gamma^{yy}\right\}},\\
&&R_{21,29}=1,R_{21,30}=1,\\
&&R_{22,17}=\frac{\gamma^{xy}\gamma^{xz}}{\sqrt{\gamma^{xx}}\left\{-(\gamma^{xy})^2+\gamma^{xx}\gamma^{yy}\right\}},R_{22,18}=\frac{\sqrt{\gamma^{xx}}\gamma^{yy}}{-(\gamma^{xy})^2+\gamma^{xx}\gamma^{yy}},\\
&&R_{22,19}=\frac{\sqrt{\gamma^{xx}}\gamma^{xy}}{-(\gamma^{xy})^2+\gamma^{xx}\gamma^{yy}},R_{22,20}=\frac{\gamma^{xy}}{(\gamma^{xy})^2-\gamma^{xx}\gamma^{yy}},\\
&&R_{22,21}=\frac{\gamma^{xy}\left\{-(\gamma^{xz})^2+\gamma^{xx}\gamma^{zz}\right\}}{\gamma^{xx}\left\{-(\gamma^{xy})^2+\gamma^{xx}\gamma^{yy}\right\}},R_{22,22}=\frac{\gamma^{xz}\left\{(\gamma^{xy})^2+\gamma^{xx}\gamma^{yy}\right\}-2\gamma^{xx}\gamma^{xy}\gamma^{yz}}{\gamma^{xx}\left\{(\gamma^{xy})^2-\gamma^{xx}\gamma^{yy}\right\}},\\
&&R_{22,23}=\frac{\gamma^{xy}\gamma^{xz}}{\sqrt{\gamma^{xx}}\left\{(\gamma^{xy})^2-\gamma^{xx}\gamma^{yy}\right\}},R_{22,24}=\frac{\sqrt{\gamma^{xx}}\gamma^{yy}}{(\gamma^{xy})^2-\gamma^{xx}\gamma^{yy}},\\
&&R_{22,25}=\frac{\sqrt{\gamma^{xx}}\gamma^{xy}}{(\gamma^{xy})^2-\gamma^{xx}\gamma^{yy}},R_{22,26}=\frac{\gamma^{xy}}{(\gamma^{xy})^2-\gamma^{xx}\gamma^{yy}},\\
&&R_{22,27}=\frac{\gamma^{xy}\left\{-(\gamma^{xz})^2+\gamma^{xx}\gamma^{zz}\right\}}{\gamma^{xx}\left\{-(\gamma^{xy})^2+\gamma^{xx}\gamma^{yy}\right\}},R_{22,28}=\frac{\gamma^{xz}\left\{(\gamma^{xy})^2+\gamma^{xx}\gamma^{yy}\right\}-2\gamma^{xx}\gamma^{xy}\gamma^{yz}}{\gamma^{xx}\left\{(\gamma^{xy})^2-\gamma^{xx}\gamma^{yy}\right\}},\\
&&R_{23,17}=\frac{1}{\sqrt{\gamma^{xx}}},R_{23,21}=-\frac{\gamma^{xz}}{\gamma^{xx}},R_{23,22}=-\frac{\gamma^{xy}}{\gamma^{xx}},\\
&&R_{23,23}=-\frac{1}{\sqrt{\gamma^{xx}}},R_{23,27}=-\frac{\gamma^{xz}}{\gamma^{xx}},R_{23,28}=-\frac{\gamma^{xy}}{\gamma^{xx}},\\
&&R_{24,17}=\frac{\sqrt{\gamma^{xx}}\gamma^{xz}}{(\gamma^{xy})^2-\gamma^{xx}\gamma^{yy}},R_{24,18}=\frac{\sqrt{\gamma^{xx}}\gamma^{xy}}{(\gamma^{xy})^2-\gamma^{xx}\gamma^{yy}},R_{24,19}=\frac{\gamma^{xx}\sqrt{\gamma^{xx}}}{(\gamma^{xy})^2-\gamma^{xx}\gamma^{yy}},\\
&&R_{24,20}=\frac{\gamma^{xx}}{-(\gamma^{xy})^2+\gamma^{xx}\gamma^{yy}},R_{24,21}=\frac{(\gamma^{xz})^2-\gamma^{xx}\gamma^{zz}}{-(\gamma^{xy})^2+\gamma^{xx}\gamma^{yy}},\\
&&R_{24,22}=\frac{2\gamma^{xy}\gamma^{xz}-2\gamma^{xx}\gamma^{yz}}{-(\gamma^{xy})^2+\gamma^{xx}\gamma^{yy}},R_{24,23}=\frac{\sqrt{\gamma^{xx}}\gamma^{xz}}{-(\gamma^{xy})^2+\gamma^{xx}\gamma^{yy}},R_{24,24}=\frac{\sqrt{\gamma^{xx}}\gamma^{xy}}{-(\gamma^{xy})^2+\gamma^{xx}\gamma^{yy}},\\
&&R_{24,25}=\frac{\gamma^{xx}\sqrt{\gamma^{xx}}}{-(\gamma^{xy})^2+\gamma^{xx}\gamma^{yy}},R_{24,26}=\frac{\gamma^{xx}}{-(\gamma^{xy})^2+\gamma^{xx}\gamma^{yy}},R_{24,27}=\frac{(\gamma^{xz})^2-\gamma^{xx}\gamma^{zz}}{-(\gamma^{xy})^2+\gamma^{xx}\gamma^{yy}},\\
&&R_{24,28}=\frac{2\gamma^{xy}\gamma^{xz}-2\gamma^{xx}\gamma^{yz}}{-(\gamma^{xy})^2+\gamma^{xx}\gamma^{yy}},\\
&&R_{25,22}=1,R_{25,28}=1,R_{26,21}=1,R_{26,27}=1,R_{27,20}=1,R_{27,26}=1,\\
&&R_{28,0}=-\frac{\gamma^{xz}}{\gamma^{xx}},R_{28,1}=-\frac{\gamma^{xy}}{\gamma^{xx}},R_{28,19}=1,R_{28,25}=1,R_{29,1}=1,R_{29,18}=1,\\
&&R_{29,24}=1,R_{30,0}=1,R_{30,17}=1,R_{30,23}=1.
\end{eqnarray*}
Non-zero elemnts of matrix $R^{-1}_{ij}$, ($0\le i,j \le 30$) are
\begin{eqnarray*}
&&R^{-1}_{0,2}=\frac{1}{2},R^{-1}_{0,13}=\frac{1}{2}\left(\frac{(\gamma^{xy})^2}{\gamma^{xx}}-\gamma^{yy}\right),R^{-1}_{0,14}=\frac{1}{2}\left(\frac{\gamma^{xy}\gamma^{xz}}{\gamma^{xx}}-\gamma^{yz}\right),\\
&&R^{-1}_{0,18}=\frac{1}{2}\left(-\frac{(\gamma^{xy}\gamma^{xy})}{\gamma^{xx}}+\gamma^{yy}\right),R^{-1}_{0,19}=\frac{1}{2}\left(-\frac{(\gamma^{xy}\gamma^{xz})}{\gamma^{xx}}+\gamma^{yz}\right),\\
&&R^{-1}_{1,1}=\frac{1}{2},R^{-1}_{1,13}=\frac{1}{2}\left(-\frac{\gamma^{xy}\gamma^{xz}}{\gamma^{xx}}+\gamma^{yz}\right),R^{-1}_{1,14}=\frac{1}{2}\left(-\frac{\gamma^{xz}\gamma^{xz}}{\gamma^{xx}}+\gamma^{zz}\right),R^{-1}_{1,18}=\frac{1}{2}\left(\frac{\gamma^{xy}\gamma^{xz}}{\gamma^{xx}}-\gamma^{yz}\right),\\
&&R^{-1}_{1,19}=\frac{1}{2}\left(\frac{(\gamma^{xz})^2}{\gamma^{xx}}-\gamma^{zz}\right),R^{-1}_{2,20}=1,R^{-1}_{3,19}=1,R^{-1}_{4,18}=1,\\
&&R^{-1}_{5,17}=1,R^{-1}_{6,16}=1,R^{-1}_{7,15}=1,R^{-1}_{8,14}=1,R^{-1}_{9,13}=1,\\
&&R^{-1}_{10,12}=1,R^{-1}_{11,11}=1,R^{-1}_{12,10}=1,R^{-1}_{13,9}=1,\\
&&R^{-1}_{14,2}=-\frac{1}{2\gamma^{xx}},R^{-1}_{14,5}=1,R^{-1}_{14,7}=\frac{\gamma^{xy}}{\gamma^{xx}},\\
&&R^{-1}_{14,8}=\frac{\gamma^{xz}}{\gamma^{xx}},R^{-1}_{14,11}=-\frac{\gamma^{xy}}{2\gamma^{xx}},R^{-1}_{14,13}=\frac{\gamma^{yy}}{2\gamma^{xx}},\\
&&R^{-1}_{14,14}=\frac{\gamma^{yz}}{2\gamma^{xx}},R^{-1}_{14,16}=-\frac{\gamma^{xy}}{2\gamma^{xx}},R^{-1}_{14,17}=-\frac{\gamma^{xz}}{\gamma^{xx}},\\
&&R^{-1}_{14,18}=-\frac{\gamma^{yy}}{2\gamma^{xx}},R^{-1}_{14,19}=-\frac{\gamma^{yz}}{2\gamma^{xx}},R^{-1}_{14,30}=\frac{1}{\gamma^{xx}},\\
&&R^{-1}_{15,1}=-\frac{1}{2\gamma^{xx}},R^{-1}_{15,4}=1,R^{-1}_{15,6}=\frac{\gamma^{xy}}{\gamma^{xx}},\\
&&R^{-1}_{15,7}=\frac{\gamma^{xz}}{\gamma^{xx}},R^{-1}_{15,10}=-\frac{\gamma^{xy}}{\gamma^{xx}},R^{-1}_{15,11}=-\frac{\gamma^{xz}}{2\gamma^{xx}},\\
&&R^{-1}_{15,13}=-\frac{\gamma^{yz}}{2\gamma^{xx}},R^{-1}_{15,14}=-\frac{\gamma^{zz}}{2\gamma^{xx}},R^{-1}_{15,16}=-\frac{\gamma^{xz}}{2\gamma^{xx}},\\
&&R^{-1}_{15,18}=\frac{\gamma^{yz}}{2\gamma^{xx}},R^{-1}_{15,19}=\frac{\gamma^{zz}}{2\gamma^{xx}},R^{-1}_{15,29}=\frac{1}{\gamma^{xx}},\\
&&R^{-1}_{16,0}=-\frac{1}{f\gamma^{xx}},R^{-1}_{16,1}=\frac{(-1+f)\gamma^{xy}}{f(\gamma^{xx})^2},R^{-1}_{16,2}=\frac{(-1+f)\gamma^{xz}}{f(\gamma^{xx})^2},\\
&&R^{-1}_{16,3}=1,R^{-1}_{16,6}=\frac{(\gamma^{xy})^2(-2+m)-\gamma^{xx}\gamma^{yy}(-1+m)}{(\gamma^{xx})^2},\\
&&R^{-1}_{16,7}=\frac{2(\gamma^{xy}\gamma^{xz}(-2+m)-\gamma^{xx}\gamma^{yz}(-1+m))}{(\gamma^{xx})^2},R^{-1}_{16,8}=\frac{(\gamma^{xz})^2(-2+m)-\gamma^{xx}\gamma^{zz}(-1+m)}{(\gamma^{xx})^2},\\
&&R^{-1}_{16,10}=\frac{(\gamma^{xy})^2(2-m)+\gamma^{xx}\gamma^{yy}(-1+m)}{(\gamma^{xx})^2},R^{-1}_{16,11}=\frac{\gamma^{xy}\gamma^{xz}(2-m)+\gamma^{xx}\gamma^{yz}(-1+m)}{(\gamma^{xx})^2},\\
&&R^{-1}_{16,13}=\frac{(\gamma^{xz}\gamma^{yy}-\gamma^{xy}\gamma^{yz})(-2+m)}{(\gamma^{xx})^2},R^{-1}_{16,14}=\frac{(\gamma^{xz}\gamma^{yz}-\gamma^{xy}\gamma^{zz})(-2+m)}{(\gamma^{xx})^2},\\
&&R^{-1}_{16,16}=\frac{\gamma^{xy}\gamma^{xz}(2-m)+\gamma^{xx}\gamma^{yz}(-1+m)}{(\gamma^{xx})^2},R^{-1}_{16,17}=\frac{(\gamma^{xz})^2(2-m)+\gamma^{xx}\gamma^{zz}(-1+m)}{(\gamma^{xx})^2},\\
&&R^{-1}_{16,18}=-\frac{(\gamma^{xz}\gamma^{yy}-\gamma^{xy}\gamma^{yz})(-2+m)}{(\gamma^{xx})^2},R^{-1}_{16,19}=-\frac{(\gamma^{xz}\gamma^{yz}-\gamma^{xy}\gamma^{zz})(-2+m)}{(\gamma^{xx})^2},\\
&&R^{-1}_{16,28}=\frac{m}{\gamma^{xx}},R^{-1}_{16,29}=\frac{\gamma^{xy}(-2+m)}{(\gamma^{xx})^2},R^{-1}_{16,30}=\frac{\gamma^{xz}(-2+m)}{(\gamma^{xx})^2},\\
&&R^{-1}_{17,2}=-\frac{1}{4},R^{-1}_{17,13}=\frac{1}{4}\left(-\frac{(\gamma^{xy})^2}{\gamma^{xx}}+\gamma^{yy}\right),\\
&&R^{-1}_{17,14}=\frac{1}{4}\left(-\frac{\gamma^{xy}\gamma^{xz}}{\gamma^{xx}}+\gamma^{yz}\right),R^{-1}_{17,18}=\frac{1}{4}\left(\frac{(\gamma^{xy})^2}{\gamma^{xx}}-\gamma^{yy}\right),\\
&&R^{-1}_{17,19}=\frac{1}{4}\left(\frac{\gamma^{xy}\gamma^{xz}}{\gamma^{xx}}-\gamma^{yz}\right),R^{-1}_{17,23}=\frac{\sqrt{\gamma^{xx}}}{2},R^{-1}_{17,25}=\frac{\gamma^{xy}}{2\sqrt{\gamma^{xx}}},\\
&&R^{-1}_{17,26}=\frac{\gamma^{xz}}{2\sqrt{\gamma^{xx}}},R^{-1}_{17,30}=\frac{1}{2},\\
&&R^{-1}_{18,1}=-\frac{1}{4},R^{-1}_{18,13}=\frac{1}{4}\left(\frac{\gamma^{xy}\gamma^{xz}}{\gamma^{xx}}-\gamma^{yz}\right),R^{-1}_{18,14}=\frac{1}{4}\left(\frac{(\gamma^{xz})^2}{\gamma^{xx}}-\gamma^{zz}\right),\\
&&R^{-1}_{18,18}=\frac{1}{4}\left(-\frac{\gamma^{xy}\gamma^{xz}}{\gamma^{xx}}+\gamma^{yz}\right),R^{-1}_{18,19}=\frac{1}{4}\left(-\frac{(\gamma^{xz})^2}{\gamma^{xx}}+\gamma^{zz}\right),\\
&&R^{-1}_{18,22}=\frac{\sqrt{\gamma^{xx}}}{2},R^{-1}_{18,24}=\frac{\gamma^{xy}}{2\sqrt{\gamma^{xx}}},R^{-1}_{18,25}=\frac{\gamma^{xz}}{2\sqrt{\gamma^{xx}}},R^{-1}_{18,29}=\frac{1}{2},\\
&&R^{-1}_{19,1}=\frac{\gamma^{xy}}{4\gamma^{xx}},R^{-1}_{19,2}=\frac{\gamma^{xz}}{4\gamma^{xx}},R^{-1}_{19,13}=\frac{-\gamma^{xz}\gamma^{yy}+\gamma^{xy}\gamma^{yz}}{4\gamma^{xx}},\\
&&R^{-1}_{19,14}=\frac{-\gamma^{xz}\gamma^{yz}+\gamma^{xy}\gamma^{zz}}{4\gamma^{xx}},R^{-1}_{19,18}=\frac{\gamma^{xz}\gamma^{yy}-\gamma^{xy}\gamma^{yz}}{4\gamma^{xx}},R^{-1}_{19,19}=\frac{\gamma^{xz}\gamma^{yz}-\gamma^{xy}\gamma^{zz}}{4\gamma^{xx}},\\
&&R^{-1}_{19,22}=-\frac{\gamma^{xy}}{2\sqrt{\gamma^{xx}}},R^{-1}_{19,23}=-\frac{\gamma^{xz}}{2\sqrt{\gamma^{xx}}},R^{-1}_{19,24}=-\frac{\gamma^{yy}}{2\sqrt{\gamma^{xx}}},\\
&&R^{-1}_{19,25}=-\frac{\gamma^{yz}}{\sqrt{\gamma^{xx}}},R^{-1}_{19,26}=-\frac{\gamma^{zz}}{2\sqrt{\gamma^{xx}}},R^{-1}_{19,27}=\frac{1}{2\sqrt{\gamma^{xx}}},R^{-1}_{19,28}=\frac{1}{2},\\
&&R^{-1}_{20,6}=\frac{(\gamma^{xy})^2-\gamma^{xx}\gamma^{yy}}{2\sqrt{\gamma^{xx}}},R^{-1}_{20,7}=\frac{\gamma^{xy}\gamma^{xz}-\gamma^{xx}\gamma^{yz}}{\sqrt{\gamma^{xx}}},R^{-1}_{20,8}=\frac{(\gamma^{xz})^2-\gamma^{xx}\gamma^{zz}}{2\sqrt{\gamma^{xx}}},\\
&&R^{-1}_{20,10}=\frac{-(\gamma^{xy})^2+\gamma^{xx}\gamma^{yy}}{2\sqrt{\gamma^{xx}}},R^{-1}_{20,11}=\frac{-(\gamma^{xy}\gamma^{xz})+\gamma^{xx}\gamma^{yz}}{2\sqrt{\gamma^{xx}}},R^{-1}_{20,13}=\frac{\gamma^{xz}\gamma^{yy}-\gamma^{xy}\gamma^{yz}}{2\sqrt{\gamma^{xx}}},\\
&&R^{-1}_{20,14}=\frac{\gamma^{xz}\gamma^{yz}-\gamma^{xy}\gamma^{zz}}{2\sqrt{\gamma^{xx}}},R^{-1}_{20,16}=\frac{-\gamma^{xy}\gamma^{xz}+\gamma^{xx}\gamma^{yz}}{2\sqrt{\gamma^{xx}}},R^{-1}_{20,17}=\frac{-(\gamma^{xz})^2+\gamma^{xx}\gamma^{zz}}{2\sqrt{\gamma^{xx}}},\\
&&R^{-1}_{20,18}=\frac{-(\gamma^{xz}\gamma^{yy})+\gamma^{xy}\gamma^{yz}}{2\sqrt{\gamma^{xx}}},R^{-1}_{20,19}=\frac{-(\gamma^{xz}\gamma^{yz})+\gamma^{xy}\gamma^{zz}}{2\sqrt{\gamma^{xx}}},R^{-1}_{20,27}=\frac{1}{2},\\
&&R^{-1}_{20,28}=\frac{\sqrt{\gamma^{xx}}}{2},R^{-1}_{20,29}=\frac{\gamma^{xy}}{2\sqrt{\gamma^{xx}}},R^{-1}_{20,30}=\frac{\gamma^{xz}}{2\sqrt{\gamma^{xx}}},\\
&&R^{-1}_{21,8}=-\frac{\sqrt{\gamma^{xx}}}{2},R^{-1}_{21,14}=-\frac{\gamma^{xy}}{2\sqrt{\gamma^{xx}}},R^{-1}_{21,17}=\frac{\sqrt{\gamma^{xx}}}{2},\\
&&R^{-1}_{21,19}=\frac{\gamma^{xy}}{2\sqrt{\gamma^{xx}}},R^{-1}_{21,26}=\frac{1}{2},R^{-1}_{22,7}=-\frac{\sqrt{\gamma^{xx}}}{2},\\
&&R^{-1}_{22,11}=\frac{\sqrt{\gamma^{xx}}}{4},R^{-1}_{22,13}=-\frac{\gamma^{xy}}{4\sqrt{\gamma^{xx}}},R^{-1}_{22,14}=\frac{\gamma^{xz}}{4\sqrt{\gamma^{xx}}},\\
&&R^{-1}_{22,16}=\frac{\sqrt{\gamma^{xx}}}{4},R^{-1}_{22,18}=\frac{\gamma^{xy}}{4\sqrt{\gamma^{xx}}},R^{-1}_{22,19}=-\frac{\gamma^{xz}}{4\sqrt{\gamma^{xx}}},R^{-1}_{22,25}=\frac{1}{2},\\
&&R^{-1}_{23,2}=-\frac{1}{4},R^{-1}_{23,13}=\frac{1}{4}\left(-\frac{(\gamma^{xy})^2}{\gamma^{xx}}+\gamma^{yy}\right),R^{-1}_{23,14}=\frac{1}{4}\left(-\frac{\gamma^{xy}\gamma^{xz}}{\gamma^{xx}}+\gamma^{yz}\right),\\
&&R^{-1}_{23,18}=\frac{1}{4}\left(\frac{(\gamma^{xy})^2}{\gamma^{xx}}-\gamma^{yy}\right),R^{-1}_{23,19}=\frac{1}{4}\left(\frac{\gamma^{xy}\gamma^{xz}}{\gamma^{xx}}-\gamma^{yz}\right),\\
&&R^{-1}_{23,23}=-\frac{\sqrt{\gamma^{xx}}}{2},R^{-1}_{23,25}=-\frac{\gamma^{xy}}{2\sqrt{\gamma^{xx}}},R^{-1}_{23,26}=-\frac{\gamma^{xz}}{2\sqrt{\gamma^{xx}}},R^{-1}_{23,30}=\frac{1}{2},\\
&&R^{-1}_{24,1}=-\frac{1}{4},R^{-1}_{24,13}=\frac{1}{4}\left(\frac{\gamma^{xy}\gamma^{xz}}{\gamma^{xx}}-\gamma^{yz}\right),R^{-1}_{24,14}=\frac{1}{4}\left(\frac{(\gamma^{xz})^2}{\gamma^{xx}}-\gamma^{zz}\right),\\
&&R^{-1}_{24,18}=\frac{1}{4}\left(-\frac{\gamma^{xy}\gamma^{xz}}{\gamma^{xx}}+\gamma^{yz}\right),R^{-1}_{24,19}=\frac{1}{4}\left(-\frac{(\gamma^{xz})^2}{\gamma^{xx}}+\gamma^{zz}\right),R^{-1}_{24,22}=-\frac{\sqrt{\gamma^{xx}}}{2},\\
&&R^{-1}_{24,24}=-\frac{\gamma^{xy}}{2\sqrt{\gamma^{xx}}},R^{-1}_{24,25}=-\frac{\gamma^{xz}}{2\sqrt{\gamma^{xx}}},R^{-1}_{24,29}=\frac{1}{2},\\
&&R^{-1}_{25,1}=\frac{\gamma^{xy}}{4\gamma^{xx}},R^{-1}_{25,2}=\frac{\gamma^{xz}}{4\gamma^{xx}},R^{-1}_{25,13}=\frac{-\gamma^{xz}\gamma^{yy}+\gamma^{xy}\gamma^{yz}}{4\gamma^{xx}},\\
&&R^{-1}_{25,14}=\frac{-\gamma^{xz}\gamma^{yz}+\gamma^{xy}\gamma^{zz}}{4\gamma^{xx}},R^{-1}_{25,18}=\frac{\gamma^{xz}\gamma^{yy}-\gamma^{xy}\gamma^{yz}}{4\gamma^{xx}},R^{-1}_{25,19}=\frac{\gamma^{xz}\gamma^{yz}-\gamma^{xy}\gamma^{zz}}{4\gamma^{xx}},\\
&&R^{-1}_{25,22}=\frac{\gamma^{xy}}{2\sqrt{\gamma^{xx}}},R^{-1}_{25,23}=\frac{\gamma^{xz}}{2\sqrt{\gamma^{xx}}},R^{-1}_{25,24}=\frac{\gamma^{yy}}{2\sqrt{\gamma^{xx}}},\\
&&R^{-1}_{25,25}=\frac{\gamma^{yz}}{\sqrt{\gamma^{xx}}},R^{-1}_{25,26}=\frac{\gamma^{zz}}{2\sqrt{\gamma^{xx}}},R^{-1}_{25,27}=-\frac{1}{2\sqrt{\gamma^{xx}}},\\
&&R^{-1}_{25,28}=\frac{1}{2},\\
&&R^{-1}_{26,6}=\frac{-(\gamma^{xy})^2+\gamma^{xx}\gamma^{yy}}{2\sqrt{\gamma^{xx}}},R^{-1}_{26,7}=\frac{-\gamma^{xy}\gamma^{xz}+\gamma^{xx}\gamma^{yz}}{\sqrt{\gamma^{xx}}},R^{-1}_{26,8}=\frac{-(\gamma^{xz})^2+\gamma^{xx}\gamma^{zz}}{2\sqrt{\gamma^{xx}}},\\
&&R^{-1}_{26,10}=\frac{(\gamma^{xy})^2-\gamma^{xx}\gamma^{yy}}{2\sqrt{\gamma^{xx}}},R^{-1}_{26,11}=\frac{\gamma^{xy}\gamma^{xz}-\gamma^{xx}\gamma^{yz}}{2\sqrt{\gamma^{xx}}},R^{-1}_{26,13}=\frac{-(\gamma^{xz}\gamma^{yy})+\gamma^{xy}\gamma^{yz}}{2\sqrt{\gamma^{xx}}},\\
&&R^{-1}_{26,14}=\frac{-\gamma^{xz}\gamma^{yz}+\gamma^{xy}\gamma^{zz}}{2\sqrt{\gamma^{xx}}},R^{-1}_{26,16}=\frac{\gamma^{xy}\gamma^{xz}-\gamma^{xx}\gamma^{yz}}{2\sqrt{\gamma^{xx}}},R^{-1}_{26,17}=\frac{(\gamma^{xz})^2-\gamma^{xx}\gamma^{zz}}{2\sqrt{\gamma^{xx}}},\\
&&R^{-1}_{26,18}=\frac{\gamma^{xz}\gamma^{yy}-\gamma^{xy}\gamma^{yz}}{2\sqrt{\gamma^{xx}}},R^{-1}_{26,19}=\frac{\gamma^{xz}\gamma^{yz}-\gamma^{xy}\gamma^{zz}}{2\sqrt{\gamma^{xx}}},R^{-1}_{26,27}=\frac{1}{2},\\
&&R^{-1}_{26,28}=-\frac{\sqrt{\gamma^{xx}}}{2},R^{-1}_{26,29}=-\frac{\gamma^{xy}}{2\sqrt{\gamma^{xx}}},R^{-1}_{26,30}=-\frac{\gamma^{xz}}{2\sqrt{\gamma^{xx}}},\\
&&R^{-1}_{27,8}=\frac{\sqrt{\gamma^{xx}}}{2},R^{-1}_{27,14}=\frac{\gamma^{xy}}{2\sqrt{\gamma^{xx}}},R^{-1}_{27,17}=-\frac{\sqrt{\gamma^{xx}}}{2},\\
&&R^{-1}_{27,19}=-\frac{\gamma^{xy}}{2\sqrt{\gamma^{xx}}},R^{-1}_{27,26}=\frac{1}{2},\\
&&R^{-1}_{28,7}=\frac{\sqrt{\gamma^{xx}}}{2},R^{-1}_{28,11}=-\frac{\sqrt{\gamma^{xx}}}{4},R^{-1}_{28,13}=\frac{\gamma^{xy}}{4\sqrt{\gamma^{xx}}},R^{-1}_{28,14}=-\frac{\gamma^{xz}}{4\sqrt{\gamma^{xx}}},\\
&&R^{-1}_{28,16}=-\frac{\sqrt{\gamma^{xx}}}{4},R^{-1}_{28,18}=-\frac{\gamma^{xy}}{4\sqrt{\gamma^{xx}}},R^{-1}_{28,19}=\frac{\gamma^{xz}}{4\sqrt{\gamma^{xx}}},R^{-1}_{28,25}=\frac{1}{2},\\
&&R^{-1}_{29,0}=-\frac{1}{2\sqrt{f\gamma^{xx}}},R^{-1}_{29,1}=-\frac{\gamma^{xy}}{2\gamma^{xx}\sqrt{f\gamma^{xx}}},R^{-1}_{29,2}=-\frac{\gamma^{xz}}{2\gamma^{xx}\sqrt{f\gamma^{xx}}},\\
&&R^{-1}_{29,6}=\frac{\sqrt{f}\left(-(\gamma^{xy})^2+\gamma^{xx}\gamma^{yy}\right)(-2+m)}{2(-1+f)\gamma^{xx}\sqrt{\gamma^{xx}}},R^{-1}_{29,7}=\frac{\sqrt{f}\left(-\gamma^{xy}\gamma^{xz}+\gamma^{xx}\gamma^{yz}\right)(-2+m)}{(-1+f)\gamma^{xx}\sqrt{\gamma^{xx}}},\\
&&R^{-1}_{29,8}=\frac{\sqrt{f}\left(-(\gamma^{xz})^2+\gamma^{xx}\gamma^{zz}\right)(-2+m)}{2(-1+f)\gamma^{xx}\sqrt{\gamma^{xx}}},\\
&&R^{-1}_{29,10}=\frac{\sqrt{f}\left((\gamma^{xy})^2-\gamma^{xx}\gamma^{yy}\right)(-2+m)}{2(-1+f)\gamma^{xx}\sqrt{\gamma^{xx}}},R^{-1}_{29,11}=\frac{\sqrt{f}\left(\gamma^{xy}\gamma^{xz}-\gamma^{xx}\gamma^{yz}\right)(-2+m)}{2(-1+f)\gamma^{xx}\sqrt{\gamma^{xx}}},\\
&&R^{-1}_{29,13}=\frac{\sqrt{f}\left(-\gamma^{xz}\gamma^{yy}+\gamma^{xy}\gamma^{yz}\right)(-2+m)}{2(-1+f)\gamma^{xx}\sqrt{\gamma^{xx}}},R^{-1}_{29,14}=\frac{\sqrt{f}\left(-\gamma^{xz}\gamma^{yz}+\gamma^{xy}\gamma^{zz}\right)(-2+m)}{2(-1+f)\gamma^{xx}\sqrt{\gamma^{xx}}},\\
&&R^{-1}_{29,16}=\frac{\sqrt{f}\left(\gamma^{xy}\gamma^{xz}-\gamma^{xx}\gamma^{yz}\right)(-2+m)}{2(-1+f)\gamma^{xx}\sqrt{\gamma^{xx}}},R^{-1}_{29,17}=\frac{\sqrt{f}\left((\gamma^{xz})^2-\gamma^{xx}\gamma^{zz}\right)(-2+m)}{2(-1+f)\gamma^{xx}\sqrt{\gamma^{xx}}},\\
&&R^{-1}_{29,18}=\frac{\sqrt{f}\left(\gamma^{xz}\gamma^{yy}-\gamma^{xy}\gamma^{yz}\right)(-2+m)}{2(-1+f)\gamma^{xx}\sqrt{\gamma^{xx}}},R^{-1}_{29,19}=\frac{\sqrt{f}\left(\gamma^{xz}\gamma^{yz}-\gamma^{xy}\gamma^{zz}\right)(-2+m)}{2(-1+f)\gamma^{xx}\sqrt{\gamma^{xx}}},\\
&&R^{-1}_{29,21}=\frac{1}{2},R^{-1}_{29,22}=\frac{\gamma^{xy}}{\gamma^{xx}},R^{-1}_{29,23}=\frac{\gamma^{xz}}{\gamma^{xx}},R^{-1}_{29,24}=\frac{\gamma^{yy}}{2\gamma^{xx}},\\
&&R^{-1}_{29,25}=\frac{\gamma^{yz}}{\gamma^{xx}},R^{-1}_{29,26}=\frac{\gamma^{zz}}{2\gamma^{xx}},R^{-1}_{29,27}=\frac{2-fm}{-2\gamma^{xx}+2f\gamma^{xx}},\\
&&R^{-1}_{29,28}=-\frac{\sqrt{f}(-2+m)}{2(-1+f)\sqrt{\gamma^{xx}}},R^{-1}_{29,29}=-\frac{\sqrt{f}\gamma^{xy}(-2+m)}{2(-1+f)\gamma^{xx}\sqrt{\gamma^{xx}}},\\
&&R^{-1}_{29,30}=-\frac{\sqrt{f}\gamma^{xz}(-2+m)}{2(-1+f)\gamma^{xx}\sqrt{\gamma^{xx}}},\\
&&R^{-1}_{30,0}=\frac{1}{2\sqrt{f}\sqrt{\gamma^{xx}}},R^{-1}_{30,1}=\frac{\gamma^{xy}}{2\sqrt{f}\gamma^{xx}\sqrt{\gamma^{xx}}},R^{-1}_{30,2}=\frac{\gamma^{xz}}{2\sqrt{f}\gamma^{xx}\sqrt{\gamma^{xx}}},\\
&&R^{-1}_{30,6}=\frac{\sqrt{f}\left((\gamma^{xy})^2-\gamma^{xx}\gamma^{yy}\right)(-2+m)}{2(-1+f)\gamma^{xx}\sqrt{\gamma^{xx}}},R^{-1}_{30,7}=\frac{\sqrt{f}\left(\gamma^{xy}\gamma^{xz}-\gamma^{xx}\gamma^{yz}\right)(-2+m)}{(-1+f)\gamma^{xx}\sqrt{\gamma^{xx}}},\\
&&R^{-1}_{30,8}=\frac{\sqrt{f}\left((\gamma^{xz})^2-\gamma^{xx}\gamma^{zz}\right)(-2+m)}{2(-1+f)\gamma^{xx}\sqrt{\gamma^{xx}}},R^{-1}_{30,10}=\frac{\sqrt{f}\left(-(\gamma^{xy})^2+\gamma^{xx}\gamma^{yy}\right)(-2+m)}{2(-1+f)\gamma^{xx}\sqrt{\gamma^{xx}}},\\
&&R^{-1}_{30,11}=\frac{\sqrt{f}\left(-(\gamma^{xy}\gamma^{xz})+\gamma^{xx}\gamma^{yz}\right)(-2+m)}{2(-1+f)\gamma^{xx}\sqrt{\gamma^{xx}}},R^{-1}_{30,13}=\frac{\sqrt{f}\left(\gamma^{xz}\gamma^{yy}-\gamma^{xy}\gamma^{yz}\right)(-2+m)}{2(-1+f)\gamma^{xx}\sqrt{\gamma^{xx}}},\\
&&R^{-1}_{30,14}=\frac{\sqrt{f}\left(\gamma^{xz}\gamma^{yz}-\gamma^{xy}\gamma^{zz}\right)(-2+m)}{2(-1+f)\gamma^{xx}\sqrt{\gamma^{xx}}},R^{-1}_{30,16}=\frac{\sqrt{f}\left(-\gamma^{xy}\gamma^{xz}+\gamma^{xx}\gamma^{yz}\right)(-2+m)}{2(-1+f)\gamma^{xx}\sqrt{\gamma^{xx}}},\\
&&R^{-1}_{30,17}=\frac{\sqrt{f}\left(-(\gamma^{xz})^2+\gamma^{xx}\gamma^{zz}\right)(-2+m)}{2(-1+f)\gamma^{xx}\sqrt{\gamma^{xx}}},R^{-1}_{30,18}=\frac{\sqrt{f}\left(-\gamma^{xz}\gamma^{yy}+\gamma^{xy}\gamma^{yz}\right)(-2+m)}{2(-1+f)\gamma^{xx}\sqrt{\gamma^{xx}}},\\
&&R^{-1}_{30,19}=\frac{\sqrt{f}\left(-\gamma^{xz}\gamma^{yz}+\gamma^{xy}\gamma^{zz}\right)(-2+m)}{2(-1+f)\gamma^{xx}\sqrt{\gamma^{xx}}},R^{-1}_{30,21}=\frac{1}{2},\\
&&R^{-1}_{30,22}=\frac{\gamma^{xy}}{\gamma^{xx}},R^{-1}_{30,23}=\frac{\gamma^{xz}}{\gamma^{xx}},R^{-1}_{30,24}=\frac{\gamma^{yy}}{2\gamma^{xx}},R^{-1}_{30,25}=\frac{\gamma^{yz}}{\gamma^{xx}},\\
&&R^{-1}_{30,26}=\frac{\gamma^{zz}}{2\gamma^{xx}},R^{-1}_{30,27}=\frac{2-fm}{-2\gamma^{xx}+2f\gamma^{xx}},R^{-1}_{30,28}=\frac{\sqrt{f}(-2+m)}{2(-1+f)\sqrt{\gamma^{xx}}},\\
&&R^{-1}_{30,29}=\frac{\sqrt{f}\gamma^{xy}(-2+m)}{2(-1+f)\gamma^{xx}\sqrt{\gamma^{xx}}},R^{-1}_{30,30}=\frac{\sqrt{f}\gamma^{xz}(-2+m)}{2(-1+f)\gamma^{xx}\sqrt{\gamma^{xx}}},
\end{eqnarray*}
Diagonal elements of $\Lambda_{ii}$ ($0\le i \le 30$) are
\begin{eqnarray*}
&&\Lambda_{0,0}=0, \Lambda_{1,1}=0, \Lambda_{2,2}=0, \Lambda_{3,3}=0, \Lambda_{4,4}=0, \Lambda_{5,5}=0,\\
&&\Lambda_{6,6}=0, \Lambda_{7,7}=0, \Lambda_{8,8}=0,\Lambda_{9,9}=0, \Lambda_{10,10}=0, \Lambda_{11,11}=0,\\
&&\Lambda_{12,12}=0, \Lambda_{13,13}=0, \Lambda_{14,14}=0, \Lambda_{15,15}=0, \Lambda_{16,16}=0,\\
&&\Lambda_{17,17}=-\alpha\sqrt{\gamma^{xx}},\Lambda_{18,18}=-\alpha\sqrt{\gamma^{xx}},\Lambda_{19,19}=-\alpha\sqrt{\gamma^{xx}},\Lambda_{20,20}=-\alpha\sqrt{\gamma^{xx}},\Lambda_{21,21}=-\alpha\sqrt{\gamma^{xx}},\\
&&\Lambda_{22,22}=-\alpha\sqrt{\gamma^{xx}},\Lambda_{23,23}=\alpha\sqrt{\gamma^{xx}},\Lambda_{24,24}=\alpha\sqrt{\gamma^{xx}},\Lambda_{25,25}=\alpha\sqrt{\gamma^{xx}},\Lambda_{26,26}=\alpha\sqrt{\gamma^{xx}},\\
&&\Lambda_{27,27}=\alpha\sqrt{\gamma^{xx}},\Lambda_{28,28}=\alpha\sqrt{\gamma^{xx}},\Lambda_{29,29}=-\alpha\sqrt{f\gamma^{xx}},\Lambda_{30,30}=\alpha\sqrt{f\gamma^{xx}},
\end{eqnarray*}

\newpage
\hspace{-1.4em}\textsf{\textbf{Figure Captions}}\\\\
FIG.1 $101 \times 101 \times 101$ numerical grids for calculation of the evolution of free or stuffed black holes.\\
FIG.2-(A) Lapse profiles along $x$-axis using the FVS method in the evolution of the free black hole for $m=0$ by $\Delta t/M=0.5$.\\
FIG.2-(B) Lapse profiles along $x$-axis using the LLF method in the evolution of the free black hole for $m=0$ by $\Delta t/M=0.5$.\\
FIG.2-(C) Lapse profiles along $x$-axis using the MLLF method in the evolution of the free black hole for $m=0$ by $\Delta t/M=0.5$.\\
FIG.3 Lapse profiles along $x$-axis using the FVS method and MMC method in the evolution of the free black hole for $m=0$ at $0\le t/M \le2$ (by $\Delta t/M=0.5$) and $t/M=14$.\\
FIG.4 $Z_x$ profiles along $x$-axis using the FVS method in the evolution of the free black hole for $m=0$ by $\Delta t/M=0.5$.\\
FIG.5 $\Theta$ profiles along $x$-axis using the FVS method in the evolution of the free black hole for $m=0$ by $\Delta t/M=0.5$.\\
FIG.6 $trK$ profiles along $x$-axis using the FVS method in the evolution of the free black hole for $m=0$ by $\Delta t/M=0.5$.\\
FIG.7 Lapse profiles along $x$-axis using the FVS method and MLLF method in the evolution of the free black hole for $m=-3$ by $\Delta t/M=0.5$.\\
FIG.8 Lapse profiles along $x$-axis using the FVS method and MLLF method in the evolution of the stuffed black hole for $m=0$ by $\Delta t/M=0.5$.
\newpage
\begin{center}
\includegraphics[width=.8\linewidth]{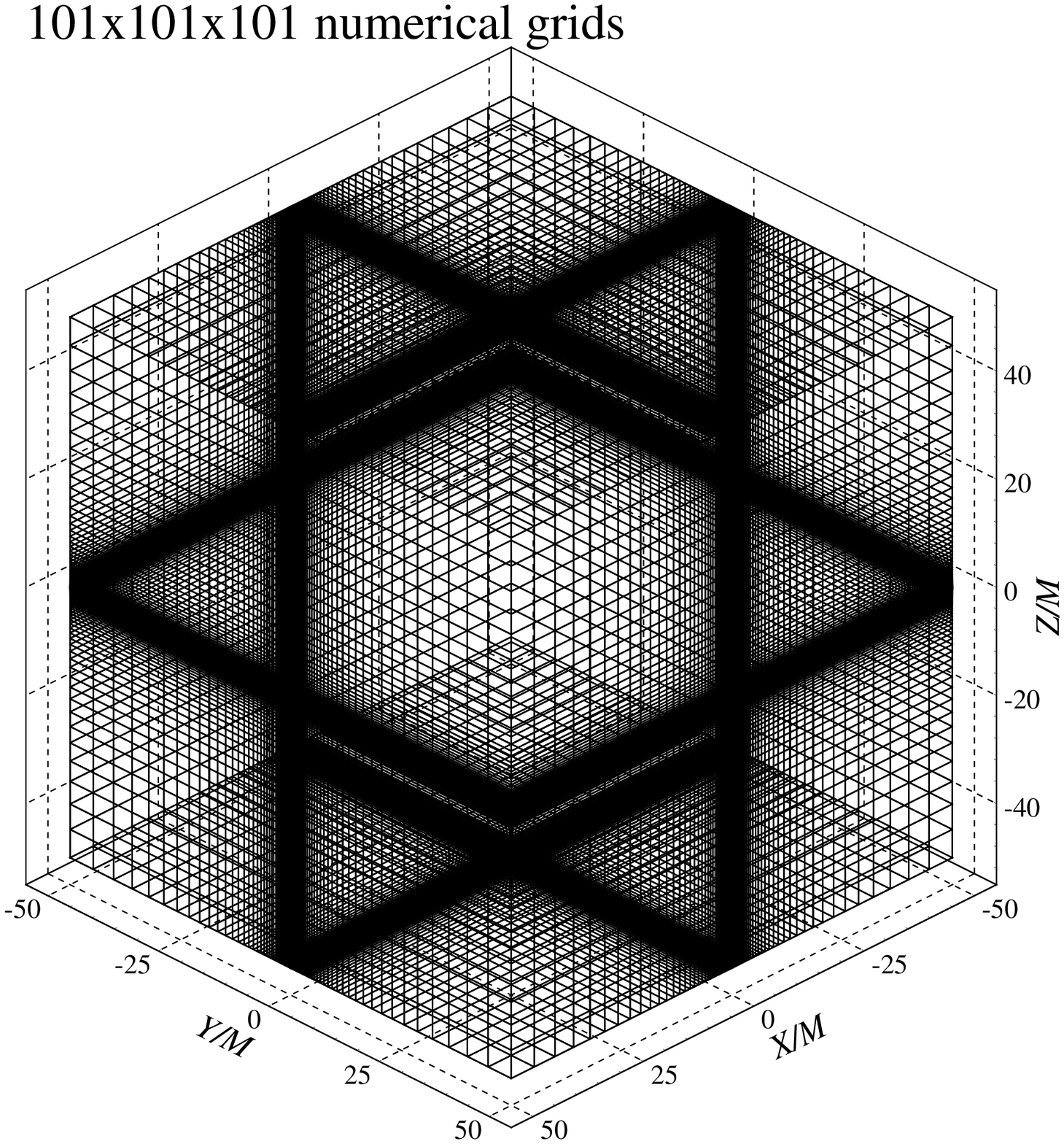}\\
\small{FIG.1 $101 \times 101 \times 101$ numerical grids for calculation of the evolution of free or stuffed black holes.}
\end{center}
\newpage
\begin{center}
\includegraphics[width=.8\linewidth]{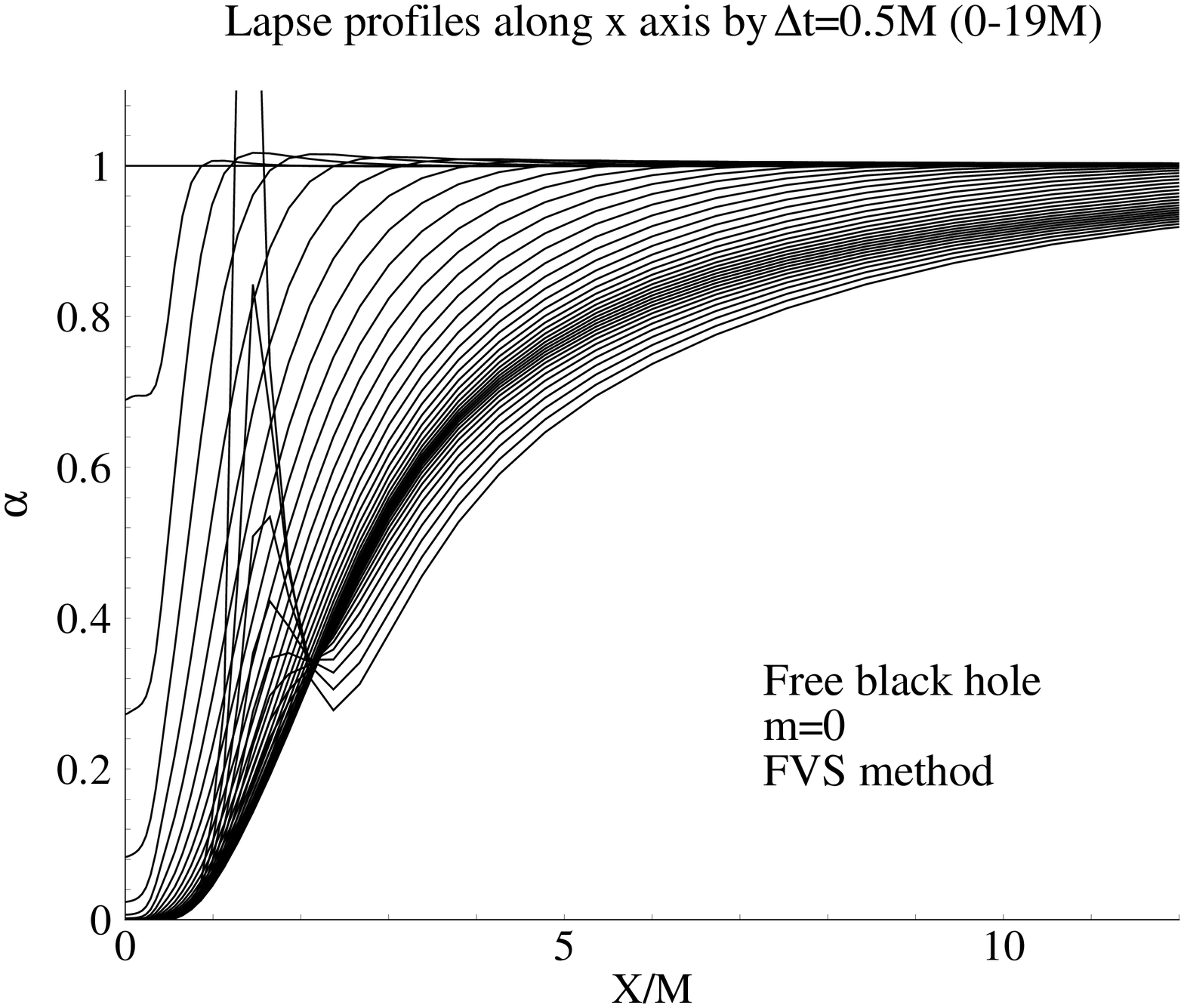}\\
\small{FIG.2-(A) Lapse profiles along $x$-axis using the FVS method in the evolution of the free black hole for $m=0$ by $\Delta t/M=0.5$.}
\end{center}
\newpage
\begin{center}
\includegraphics[width=.8\linewidth]{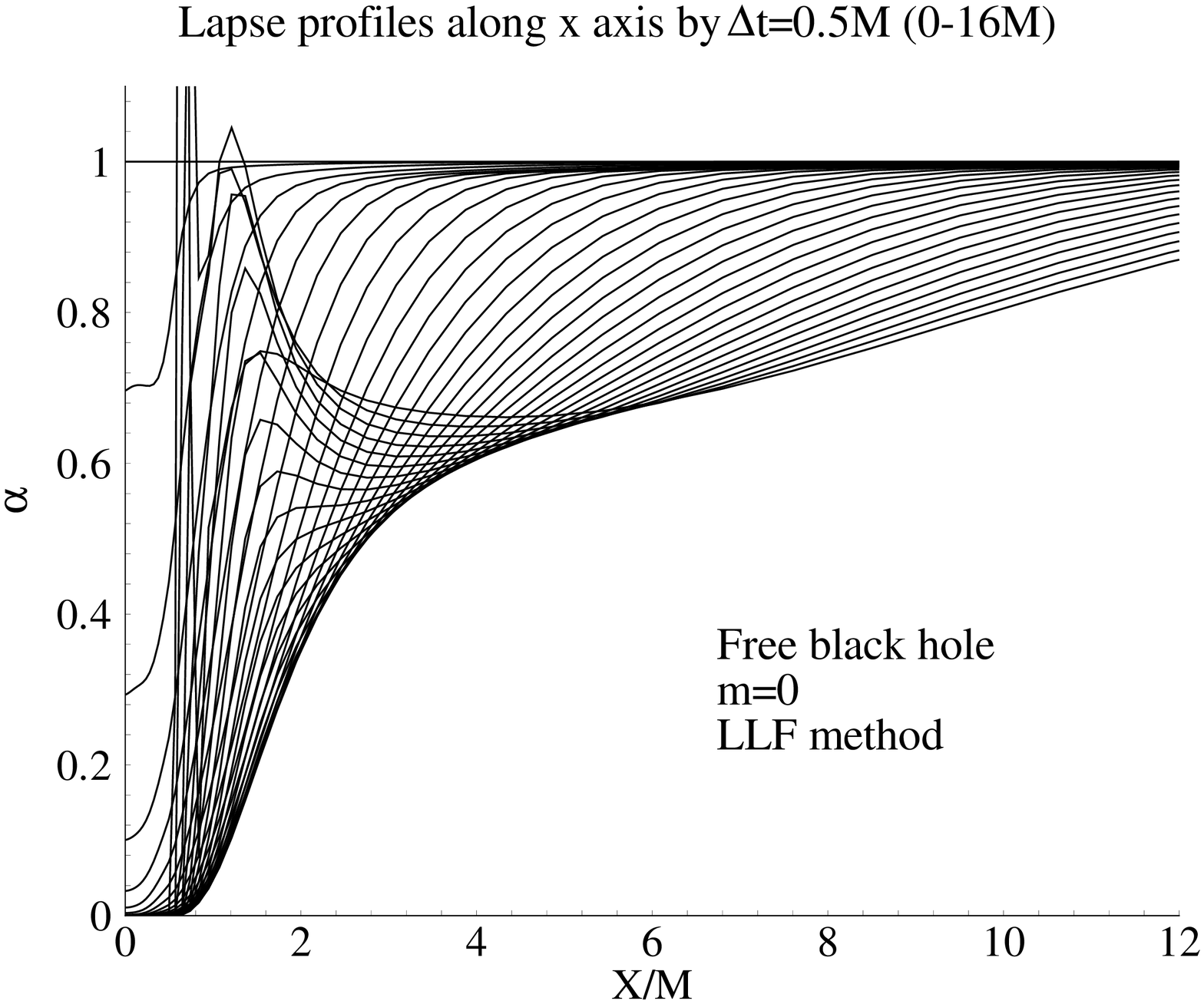}\\
\small{FIG.2-(B) Lapse profiles along $x$-axis using the LLF method in the evolution of the free black hole for $m=0$ by $\Delta t/M=0.5$.}
\end{center}
\newpage 
\begin{center}
\includegraphics[width=.8\linewidth]{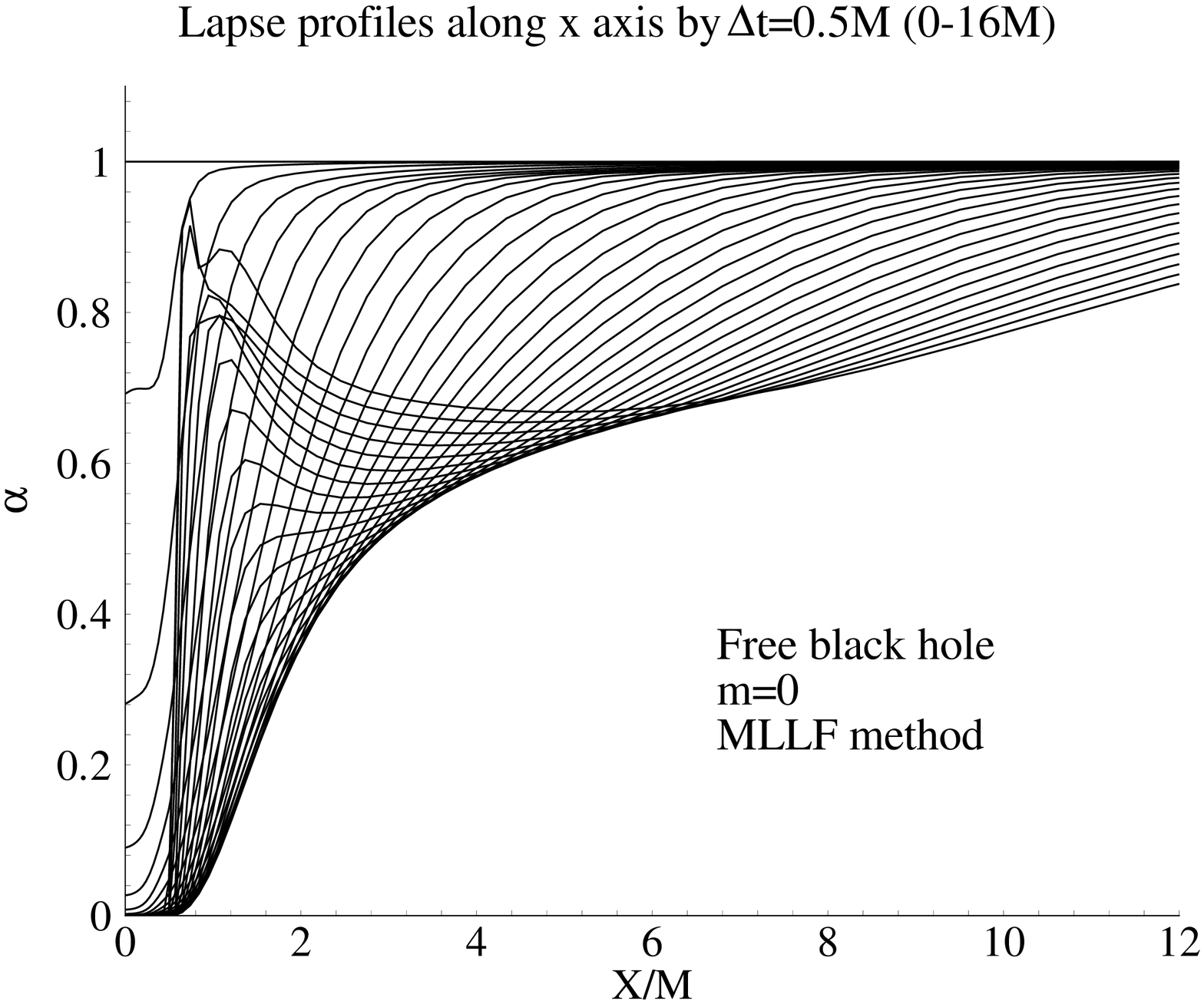}\\
\small{FIG.2-(C) Lapse profiles along $x$-axis using the MLLF method in the evolution of the free black hole for $m=0$ by $\Delta t/M=0.5$.}
\end{center}
\newpage
\begin{center}
\includegraphics[width=.8\linewidth]{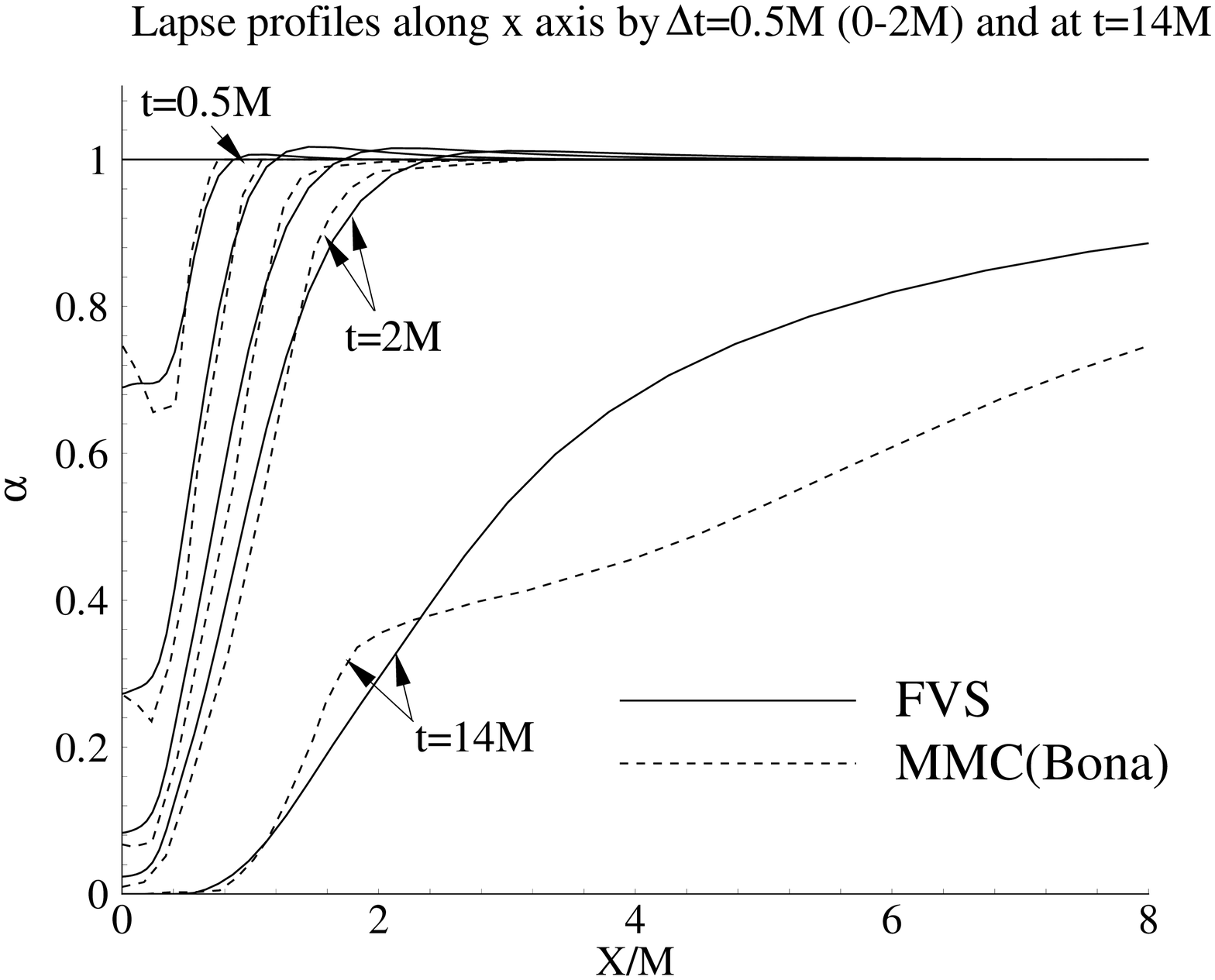}\\
\small{FIG.3  Lapse profiles along $x$-axis using the FVS method and MMC method in the evolution of the free black hole for $m=0$ at $0\le t/M \le2$ (by $\Delta t/M=0.5$) and $t/M=14$.}
\end{center}
\newpage
\begin{center}
\includegraphics[width=.8\linewidth]{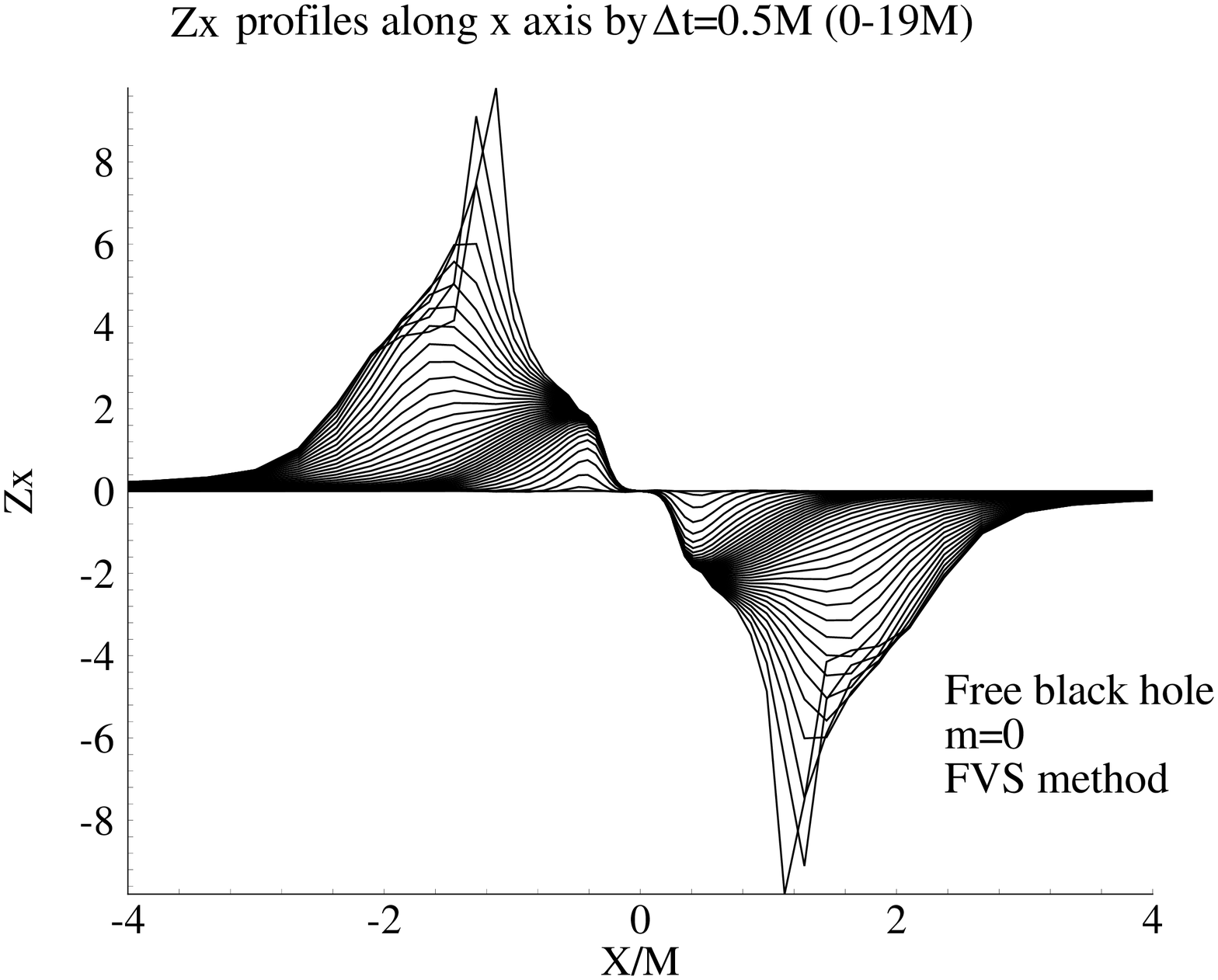}\\
\small{FIG.4 $Z_x$ profiles along $x$-axis using the FVS method in the evolution of the free black hole for $m=0$ by $\Delta t/M=0.5$.}
\end{center}
\newpage
\begin{center}
\includegraphics[width=.8\linewidth]{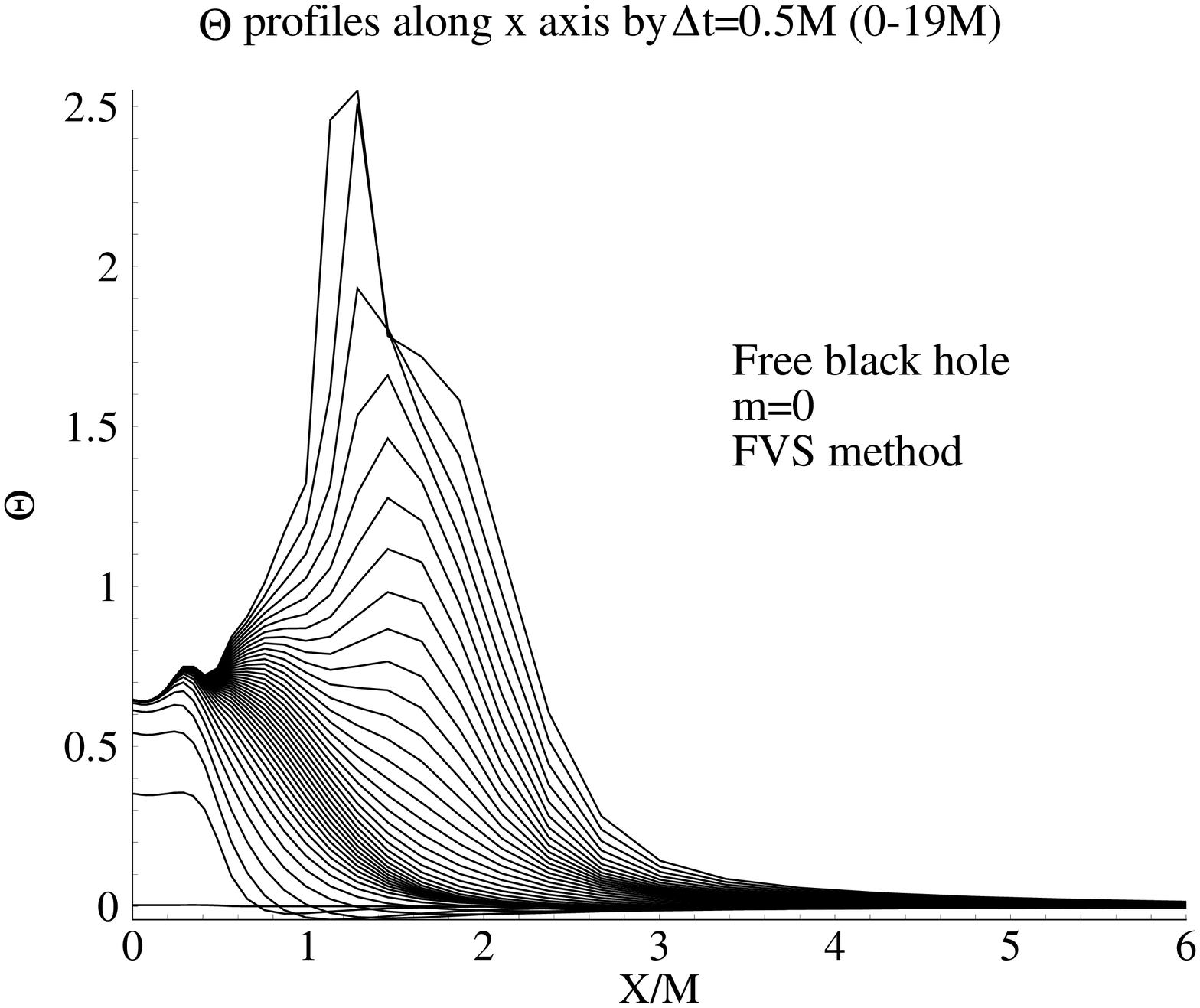}\\
\small{FIG.5 $\Theta$ profiles along $x$-axis using the FVS method in the evolution of the free black hole for $m=0$ by $\Delta t/M=0.5$.}
\end{center}
\newpage
\begin{center}
\includegraphics[width=.8\linewidth]{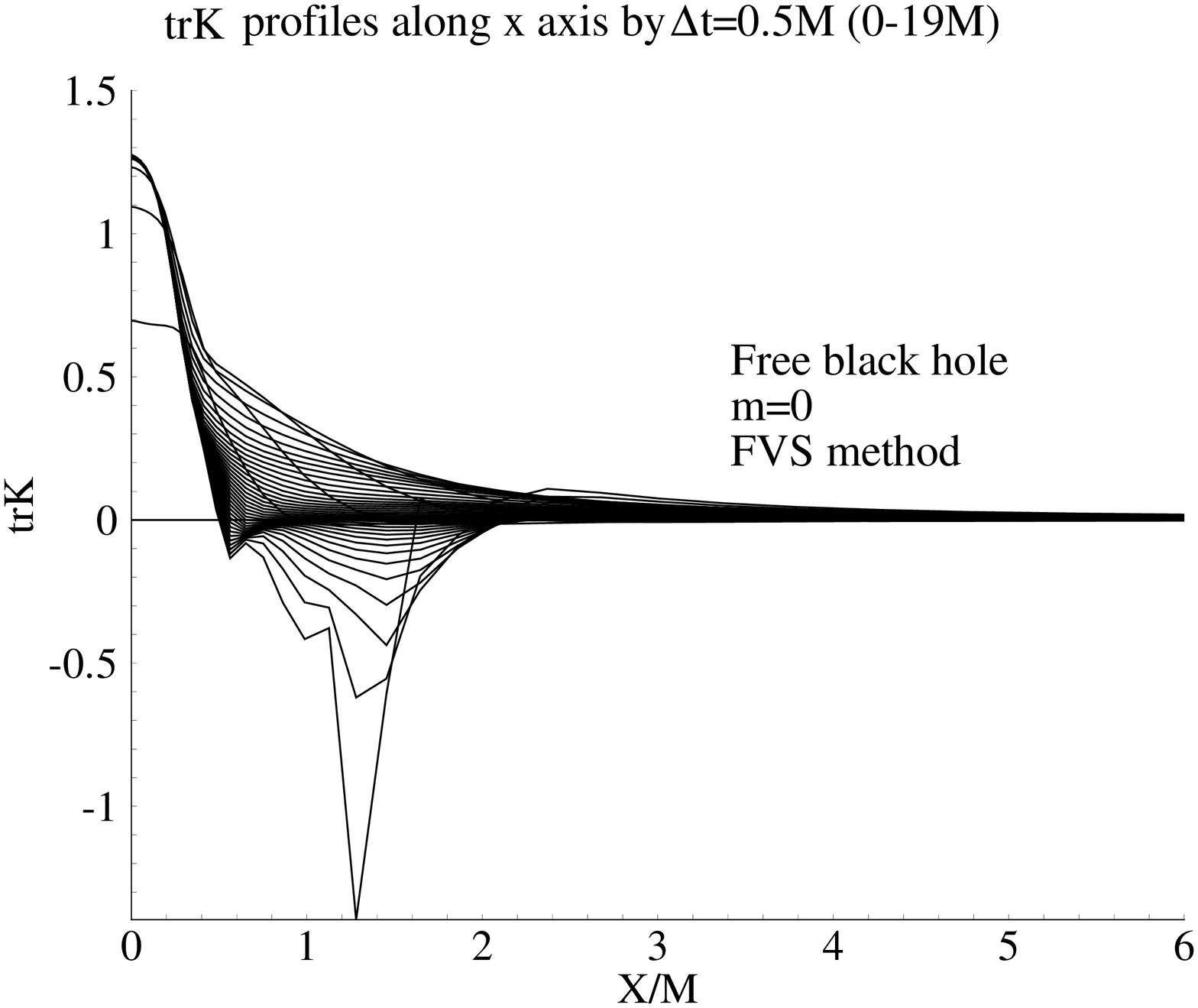}\\
\small{FIG.6 $trK$ profiles along $x$-axis using the FVS method in the evolution of the free black hole for $m=0$ by $\Delta t/M=0.5$.}
\end{center}
\newpage
\begin{center}
\includegraphics[width=.8\linewidth]{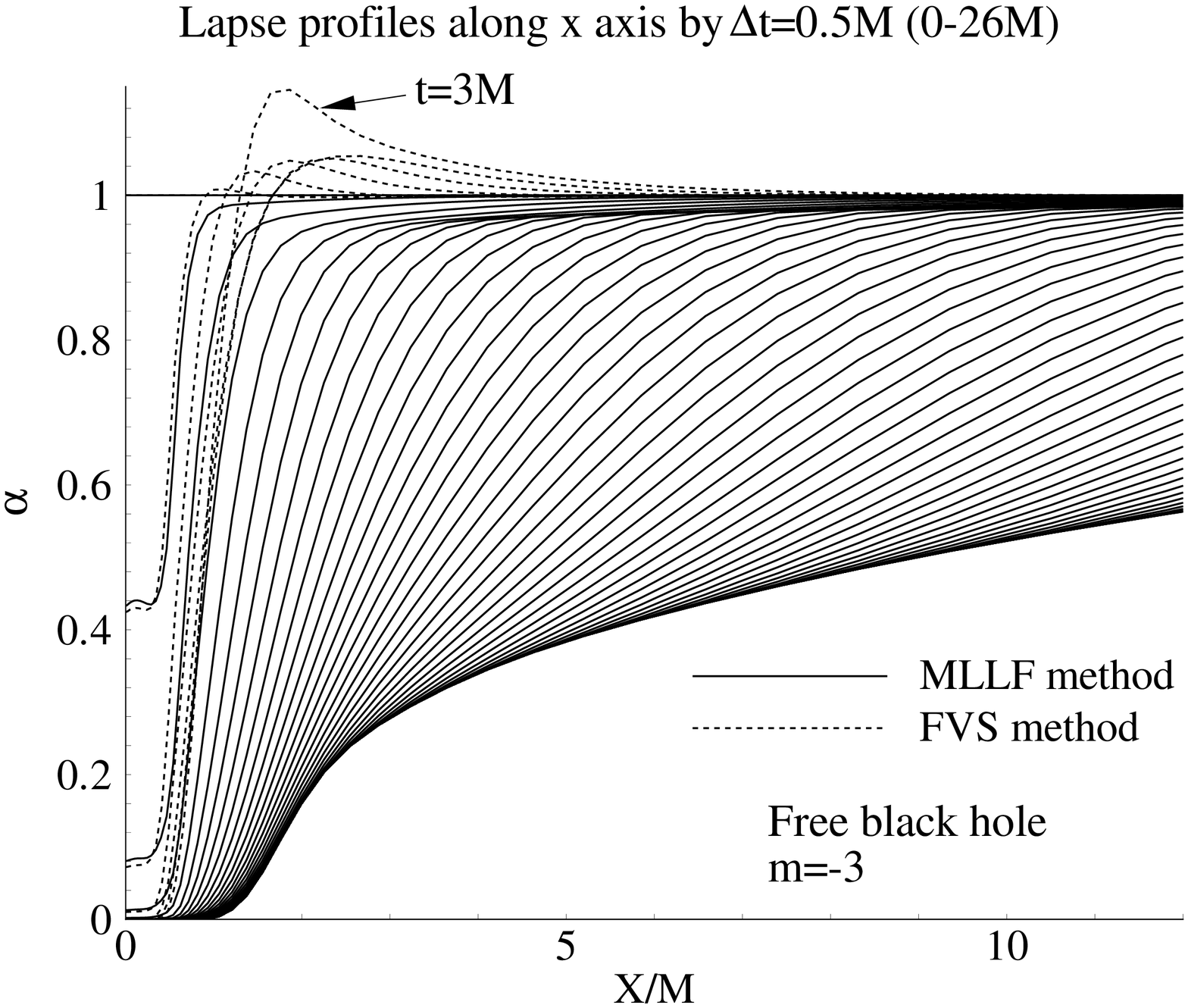}\\
\small{FIG.7 Lapse profiles along $x$-axis using the FVS method and MLLF method in the evolution of the free black hole for $m=-3$ by $\Delta t/M=0.5$.}
\end{center}
\newpage
\begin{center}
\includegraphics[width=.8\linewidth]{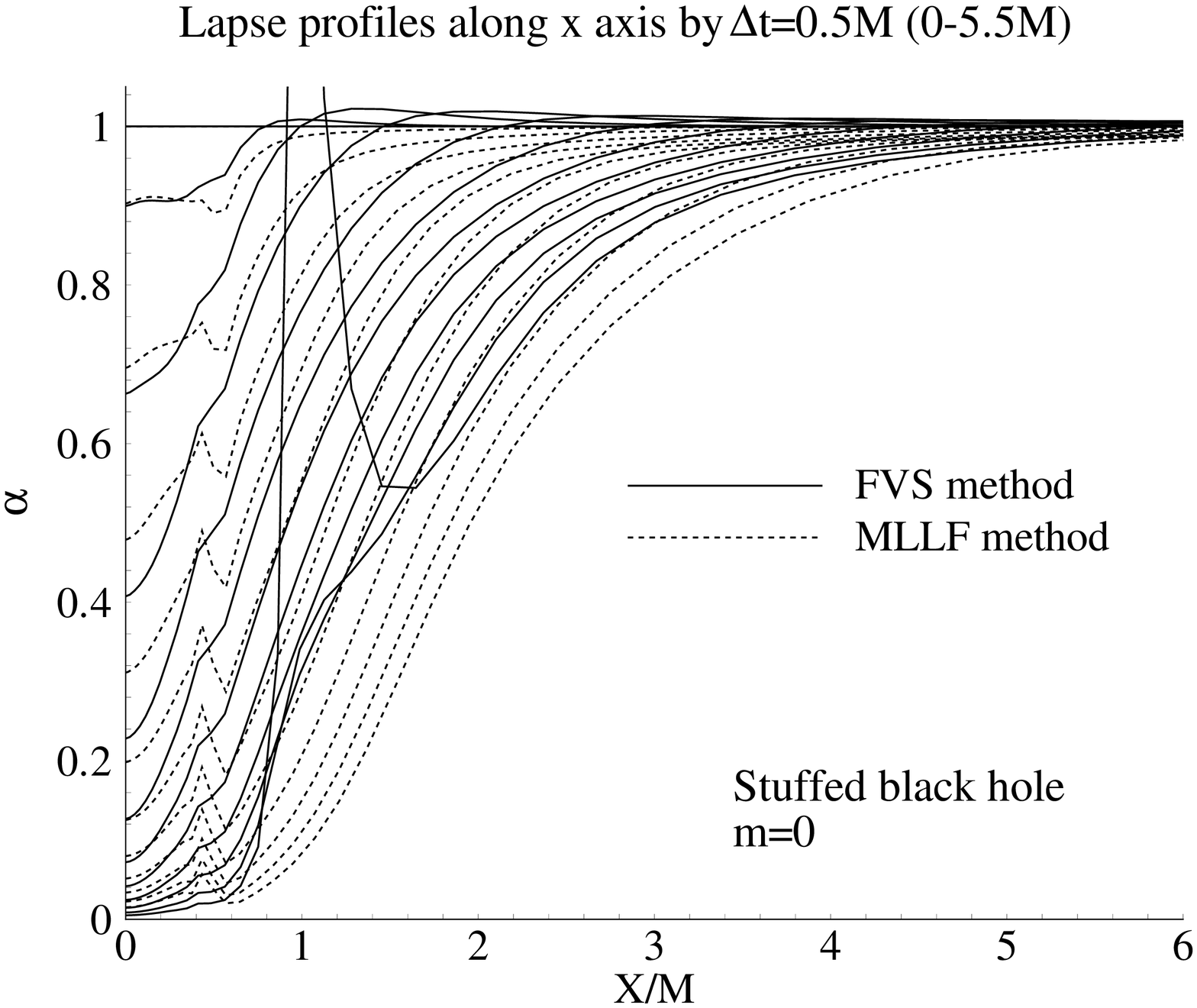}\\
\small{FIG.8 Lapse profiles along $x$-axis using the FVS method and MLLF method in the evolution of the stuffed black hole for $m=0$ by $\Delta t/M=0.5$.}
\end{center}
\end{document}